\documentclass[10pt,prd,twocolumn,floatfix,tightenlines,showpacs,showkeys,preprintnumbers,nofootinbib, longbibliography]{revtex4-1}

\pdfoutput=1
\usepackage[utf8]{inputenc}
\usepackage[colorlinks=true,breaklinks=true]{hyperref}
\hypersetup{allcolors=[rgb]{0.0 0.0 0.6},linkcolor=[rgb]{0.8 0.0 0.4}}
\usepackage{amsmath,amssymb,mathtools}
\usepackage{amsbsy}
\usepackage{epsfig}  
\usepackage{graphicx}   
\usepackage{commath}
\usepackage{slashed}             
\usepackage{url}
\usepackage{color}
\usepackage{multirow}
\usepackage{placeins}
\usepackage{hhline}
\usepackage{array}
\newcolumntype{P}[1]{>{\centering\arraybackslash}p{#1}}
\newcolumntype{M}[1]{>{\centering\arraybackslash}m{#1}}

\usepackage[dvipsnames]{xcolor}
\allowdisplaybreaks

\bibpunct{[}{]}{,}{n}{}{,}

\newcommand{\ad}[1]{{\color{black}#1}}
\newcommand{\ie}{i.\,e.}

\newcommand{\eg}{e.\,g.}

\newcommand{\sch}{Schr\"odinger}

\newcommand{\kev}{\,\,\mathrm{keV}}
\newcommand{\mev}{\,\,\mathrm{MeV}}
\newcommand{\gev}{\,\,\mathrm{GeV}}
\newcommand{\tev}{\,\,\mathrm{TeV}}
\newcommand{\cmg}{\,\mathrm{cm^2\,g^{-1}}}
\newcommand{\cms}{\,\mathrm{cm^3\,s^{-1}}}
\newcommand{\kpc}{\,\,\mathrm{kpc}}
\newcommand{\mchi}{m_\chi}
\newcommand{\mrho}{m_\rho}
\newcommand{\meta}{m_\eta}
\newcommand{\veff}{V_\text{eff}}

\newcommand{\rvec}{\mathbf{r}}
\newcommand{\pvec}{\mathbf{p}}
\newcommand{\vvec}{\mathbf{v}}
\newcommand{\sigmavec}{\pmb{\sigma}}

\newcommand{\Psil}{\Psi_\ell}

\newcommand{\ul}{u_\ell}

\newcommand{\spec}{^{2s+1}\ell_J}
\newcommand{\tdag}{T^\dagger}

\begin{document}

\title{Galactic Positron Excess from Selectively Enhanced Dark Matter Annihilation}

\author{Anirban~Das}
\email{anirbandas@theory.tifr.res.in }              
\author{Basudeb~Dasgupta}
\email{bdasgupta@theory.tifr.res.in}
\author{Anupam Ray}
\email{anupam.ray@theory.tifr.res.in}
\affiliation{Tata Institute of Fundamental Research,
             Homi Bhabha Road, Mumbai, 400005, India.}

\preprint{TIFR/TH/19-25}
\date{January 16, 2020}
\begin{abstract} 
Precision measurements of the positron flux in cosmic ray have revealed an unexplained bump in the spectrum around $E\simeq 300\,\mathrm{GeV}$, not clearly attributable to known astrophysical processes. We propose annihilation of dark matter of mass $m_\chi = 780\,\mathrm{GeV}$ with a late-time cross section $\sigma v = 4.63\times 10^{-24}\,\mathrm{cm^3\,s^{-1}}$ as a possible source. The nonmonotonic dependence of the annihilation rate on dark matter velocity, owing to a selective $p$-wave Sommerfeld enhancement, allows such a large signal from the Milky Way without violating corresponding constraints from CMB and dwarf galaxy observations. We briefly explore other signatures of this scenario, and outline avenues to test it in future experiments.
\end{abstract}

\maketitle

\section{Introduction}
Observations of the positron flux in cosmic rays have seen tremendous progress and excitement in recent years. PAMELA\,\cite{Adriani:2013uda}, Fermi-LAT\,\cite{FermiLAT:2011ab}, MASS\,\cite{Grimani:2002yz}, \mbox{Wizard/CAPRICE}\,\cite{Boezio:2001dtm}, AMS-01\,\cite{Aguilar:2007yf}, and HEAT\,\cite{DuVernois:2001bb} have reported the positron flux in the cosmic ray, with PAMELA and Fermi-LAT providing evidence for a rising positron flux around $100\gev$. The AMS-02 collaboration, in its latest published results with three times more statistics than previous measurements, has presented the positron flux up to energy $1\tev$, and the existence of a bump in the flux spectrum at $E \simeq 300\gev$ followed by a cutoff at $E \simeq 810\gev$ has been confirmed\,\cite{Aguilar:2019owu}.

\begin{figure}[t]
\begin{center}
  \includegraphics[width=0.93\columnwidth]{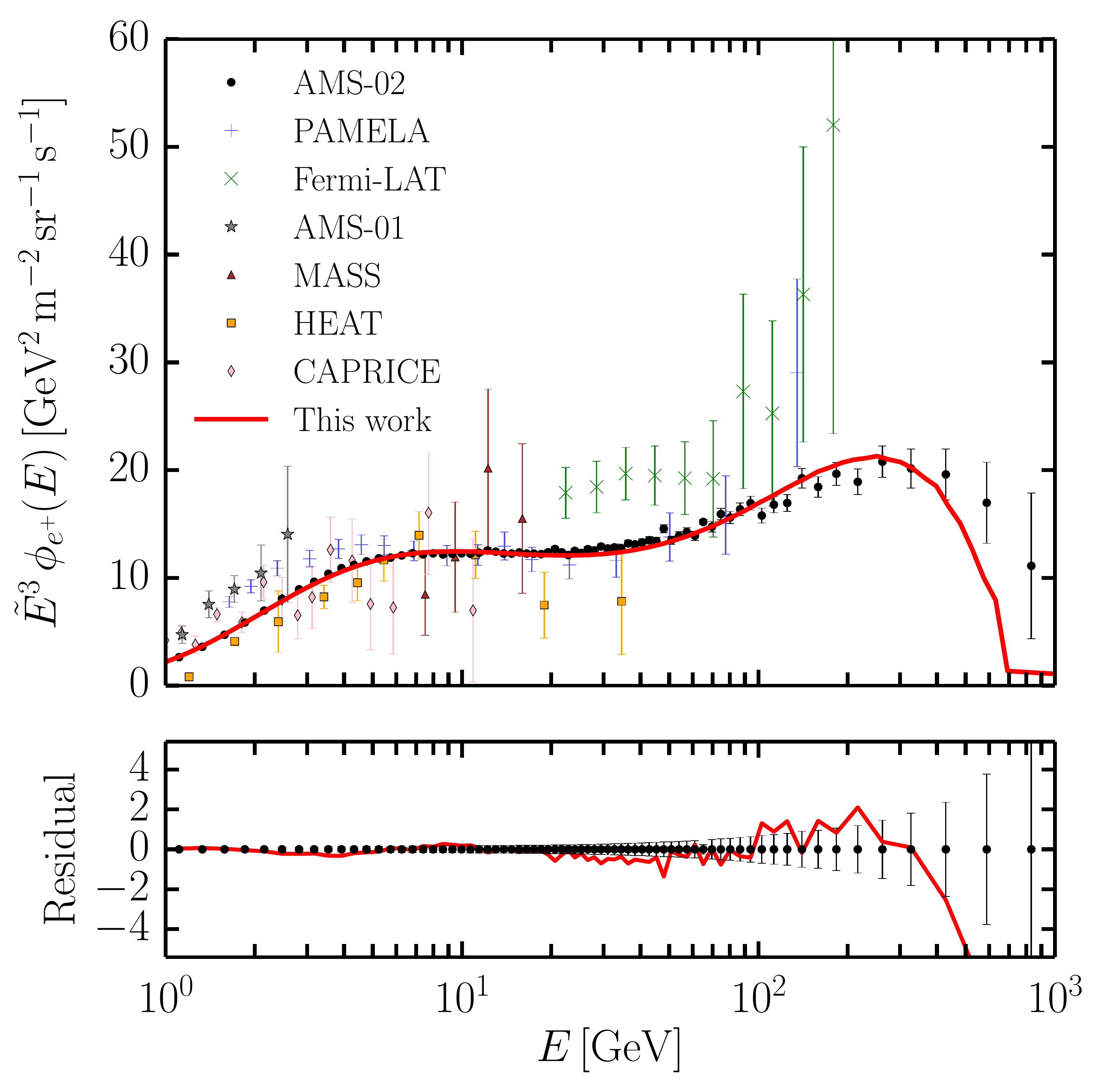}
 \caption{AMS-02 positron flux measurements (black dots), previous data (other symbols), and our theoretical prediction (red solid curve) due to Sommerfeld enhanced $p$-wave DM annihilation rate $(\sigma v)_p =$ \mbox{$4.63\times 10^{-24}\cms$} and DM mass $\mchi = 780\gev$. \ad{The red curve includes positrons from both the astrophysical background and DM annihilation.} The theoretical residuals over the AMS-02 data are shown in the lower panel. \ad{The goodness of fit $\chi^2/\mathrm{d.o.f.}=2.07$.}}\label{fig:positronflux1}
\end{center}
\end{figure}

The positron flux spectrum shows several distinct features. The flattening at $E \simeq 8\gev$ seen in Fig.\,\ref{fig:positronflux1}, is understood to be mainly caused by the diffusion of the positrons produced from the decay of the scattering products of cosmic ray with the interstellar gas\,\cite{Trotta_2011,2010A&A...524A..51D}, as described in Ref.\,\cite{Aguilar:2019owu}. However, the shape of the spectrum in the region $E > 100\gev$, a rise followed by a sharp drop, is not yet fully understood. 

DM annihilation/decay and particle acceleration by pulsars have been shown to be able to reproduce the positron spectrum but several concerns remain\,\cite{Hooper:2008kg,Yuan:2013eja,Cholis:2013psa,Ibarra:2013zia,Nardi:2008ix,Farzan:2019qdm,Profumo:2019pob}. DM of mass $(1.5-3)\tev$ and annihilation cross section $(6-23)\times 10^{-24}\cms$\,\cite{Yuan:2013eja,Cholis:2013psa}, such that it first annihilates into light intermediate states which subsequently decay to charged leptons, can produce a good fit to the earlier AMS data\,\cite{PhysRevLett.110.141102}. However, this large annihilation cross section into leptons is disallowed by the Fermi-LAT observations of the dwarf galaxies\,\cite{Ackermann:2015zua}, which show no excess gamma ray flux. Such large DM annihilation cross section into visible sector particles during the recombination era is also highly constrained by CMB observation\,\cite{Aghanim:2018eyx,Liu:2016cnk,Kawasaki:2015peu,Galli:2011rz,Slatyer:2009yq}. Moreover, note that this cross section is approximately two orders of magnitude larger than the canonical value, $(2-3)\times 10^{-26}\cms$, needed to produce the correct relic abundance of thermal DM\,\cite{Steigman:2012nb}. The possibility of a boost in the annihilation due to the substructures in the DM halo was also explored\,\cite{Hooper:2003ad}, but such large boosts are unlikely due to substructures alone. Alternatively, the pulsars in the Milky Way (MW) have also been used to explain the positron spectrum with a possibility that the nearby young pulsars (\eg, Geminga and B0656+14) contributing the most\,\cite{Hooper:2008kg,Yuan:2013eja,Cholis:2013psa,Hooper:2017gtd}. \ad{Authors in Ref.\,\cite{Hooper:2017gtd} used the HAWC data to show that these nearby pulsars are able to produce high-energy electrons, and could source of the local positron flux.} However  more recently, Ref.\,\cite{Shao-Qiang:2018zla} used the HAWC and Fermi-LAT data to show that the diffusion of the positrons from the pulsars Geminga and B0656+14 is less efficient in the energy range $(50-500)\gev$ in a two-zone diffusion model of cosmic ray, but can contribute a substantial fraction of the positron flux. Therefore, the pulsar origin of the positron flux is also debated. All things considered, the shape of the positron flux spectrum in the high energy region still remains poorly understood.

\begin{figure}[t]
\begin{center}
  \includegraphics[width=0.95\columnwidth]{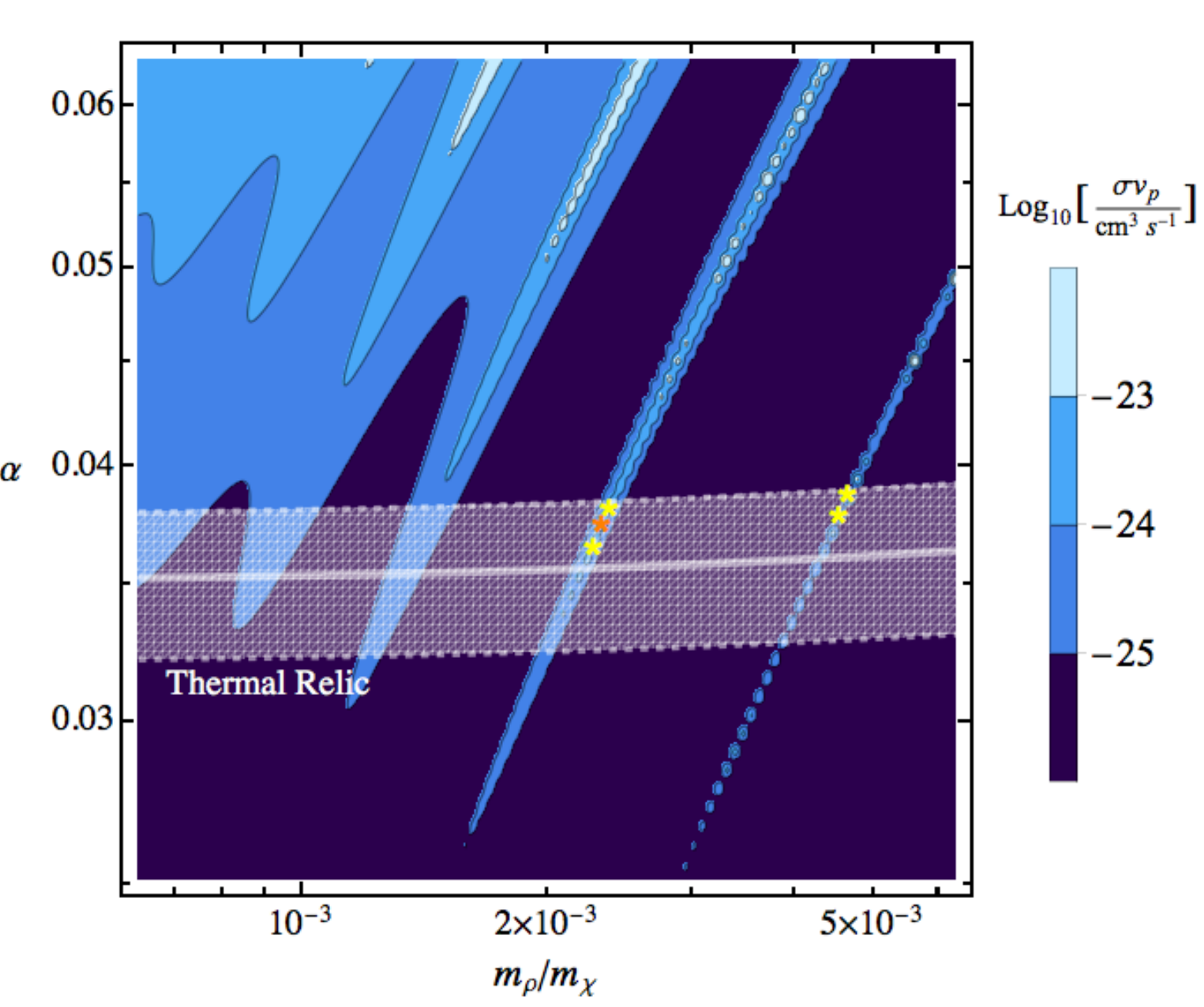}
 \caption{A contour plot showing the DM annihilation rate in the Milky Way galaxy in the $\mrho/\mchi-\alpha$ plane. Overlaid is a white band of the region of parameter space that yields correct thermal relic abundance. The thick white line denotes the contour of $\sigma v_\mathrm{relic}=2.5\times 10^{-26}\cms$, bordered by the dashed contours of $\sigma v_\mathrm{relic}=3\times 10^{-26}\cms$ (above) and $2\times 10^{-26}\cms$ (below). The points marked with asterisks can provide large enough annihilation cross section to explain the positron flux excess. In this work, we use the point marked with orange asterisk.}\label{fig:contour}
\end{center}
\end{figure}


In this work, we show that it is possible to explain the positron excess using \emph{selectively enhanced, velocity-dependent $p$-wave annihilation of Majorana DM particles}, following our earlier work\,\cite{Das:2016ced}. An angular momentum and spin-dependent selection rule in Sommerfeld effect enhances the $p$-wave DM annihilation. The $p$-wave nature implies that the rate is a nonmonotonic function of DM velocity. With a suitable choice of parameters, it can be maximum in the Milky Way-like galaxies, and less in smaller dwarf galaxies. The DM annihilation rate in the MW is shown in the contour plot in Fig.\,\ref{fig:contour}. In this parameter space, there are a few islands, marked with yellow asterisks, that yield such large galactic annihilation signal, and satisfy the thermal relic constraint, \ie, \mbox{$\langle\sigma v\rangle_\mathrm{relic} \simeq (2-3)\times 10^{-26}\cms$}\,\cite{Steigman:2012nb}, the dwarf galaxy bound from Fermi-LAT\,\cite{Ackermann:2015zua}, and DM annihilation constraint from CMB observation\,\cite{Aghanim:2018eyx}, all at the same time. A representative positron flux spectrum is shown in Fig.\,\ref{fig:positronflux1}. In this scenario, Majorana DM particles of mass \mbox{$\mchi = 780\gev$} annihilate into light dark sector scalars which promptly decay into electrons and muons. The annihilation rate in the Milky Way is \mbox{$\sigma v \simeq 4.63\times 10^{-24}\cms$}. Our purpose, in the remaining part of the paper, is to provide some details of this result and the related technical background, and to outline other phenomenological consequences of this model.

The paper is structured as follows: In Section\,\ref{sec:model}, we describe the dark sector of the model and its connection to the Standard Model (SM), followed by Section\,\ref{sec:history} discussing about the cosmological thermal history. In Section\,\ref{sec:annihilation}, we compute and explain the selective Sommerfeld enhancement in annihilation, and in Section\,\ref{sec:pheno} outline the relevant phenomenology. We summarize and conclude in Section\,\ref{sec:conclusions}.

\section{The Model}\label{sec:model}
We extend the SM by a dark sector that has a complex scalar $\Phi$ and a Dirac fermion $\chi$, with charges $-2$ and $+1$, respectively, under an approximate global $U(1)$ symmetry. The right-handed leptons in the Standard Model, $l_R$, are also charged $+2$ under this symmetry. The SM Lagrangian is therefore extended by $\mathcal{L}_{\rm BSM} \supset  \mathcal{L}_{\rm dark} + \mathcal{L}_\mathrm{portal}$, where
  \begin{equation}\label{eq:lag}
 \begin{array}{rcl}
\mathcal{L}_{\rm dark} &=& \partial^\mu \Phi^\dagger \partial_\mu \Phi + \mu^2 |\Phi|^2 - \lambda |\Phi|^4\\[1ex]  && + i \overline{\chi} \slashed{\partial}\chi - M \overline{\chi}\chi \\[1ex] && - \left({f}\Phi \chi^T \mathcal{C} \chi /{\sqrt{2}} + \text{h.c.} \right)\quad{\rm and}\\[1ex]
 \mathcal{L}_\mathrm{portal} &=& \displaystyle \sum_{l=e,\mu} \dfrac{c_l}{\Lambda}\Phi  H\bar{l}_Ll_R + \text{h.c.}\,.
 \end{array}
 \end{equation}
The dark complex scalar $\Phi$ and the dark fermion $\chi$ are coupled to each other through a Majorana-type interaction\,\cite{Weinberg:2013kea} and the 5-dimensional nonrenormalizable operator, involving the SM Higgs doublet $H$, acts as a portal between the dark and the visible sectors.  After the spontaneous breaking of the $U(1)$ and the EW symmetries, it would lead to decay of the dark scalars into electrons and muons. 

Note that the $U(1)$ is also explicitly broken at the tree-level by the electron and muon masses. It is also possible to include explicit $U(1)$-breaking terms in $\mathcal{L}_{\rm dark}$ itself. Such terms are crucial for determining the mass of the pseudo-Goldstone mode of $\Phi$, but we will not model-build that aspect and treat the pseudo-Goldstone mass as a free parameter.


  The scalar potential induces nonzero vacuum expectation value $v_\Phi$ and splits $\Phi$ into a radial mode $\rho$ and a pseudo-Goldstone mode $\eta$: $\Phi \to \left(v_\Phi+\rho+i\eta\right)/\sqrt{2}$, spontaneously breaking the $U(1)$ symmetry. 
 Due to the Majorana-type coupling with the scalar, $\chi$ also splits into two Majorana particles, $\chi_1 = (\chi - \chi^c)/\sqrt{2}$ and $\chi_2= (\chi + \chi^c)/\sqrt{2}$, with masses $\mchi$ and $\mchi+\Delta$ where $\mchi\equiv M-fv_\Phi$ and $\Delta\equiv 2fv_\Phi$. 
After dark symmetry-breaking, the resultant interactions between the fermions and the scalars are
 \begin{equation}\label{eq:int}
 -\frac{f}{2} \rho (\overline{\chi}_1 \chi_1 - \overline{\chi}_2 \chi_2) -\frac{f}{2} \eta (\overline{\chi}_2 \chi_1 + \overline{\chi}_1 \chi_2)\,.
 \end{equation}
 With these interactions, $\{\chi_1, \chi_2\}$ constitute a two-level inelastic self-interacting dark matter (SIDM). \ad{The stable $\chi_1$ is the DM candidate. After $\chi_1, \chi_2$ fall out of chemical equilibrium from the thermal bath, $\chi_2$ quickly decays $\chi_2 \to \chi_1 \eta$, and $\chi_1$ forms the DM abundance.} While the $\rho$ couples similar DM states, $\eta$ provides an off-diagonal interaction and couples different DM states. The symmetry-broken scalar potential reads
 \begin{equation}
  \label{eq:scalar}
  \begin{array}{rcl}
  V(\rho,\eta) &=& \dfrac{1}{2}\mrho^2\rho^2+\dfrac{1}{2}\meta^2\eta^2\\[1ex] &&+\lambda v_\Phi\left(\rho^3+\rho\eta^2\right)+\dfrac{\lambda}{4}\left(\rho^4+\eta^4+2\rho^2\eta^2\right)\,,
  \end{array}
 \end{equation}
 where $\mrho=\sqrt{2\lambda}v_\Phi$ and the mass of $\eta$ arises from the $U(1)$-breaking terms and its interaction with other particles in the thermal bath. 
 
 The higher-dimensional part of the Lagrangian yields
 \begin{equation}
 \begin{array}{rcl}
 \mathcal{L}_\mathrm{portal}&=&\displaystyle\sum_{l=e,\mu}\dfrac{c_l}{2\Lambda}\bar{l}_Ll_R\big(v_H\rho+iv_H\eta+\\[1.7ex]
   &&\quad+\rho h+i\eta h+v_\Phi h+v_\Phi v_H\big)\,.
 \end{array}
 \end{equation}
These terms allow the dark scalars $\rho$ and $\eta$ to decay into a pair of leptons. Further, the Higgs boson $h$ develops new decay channels $h\to\rho\, l^+ l^-$ and $h\to\eta\,l^+l^-$ and a new contribution to its decay into lepton pairs. The masses of the $e$ and $\mu$ get a contribution from the dark symmetry breaking as well.

At low energy, the DM annihilation phenomenology is controlled by the five free parameters $\{\mchi, ~\mrho, ~\meta, ~\alpha, ~\Delta\}$ \ad{where $\alpha \equiv f^2/(4\pi)$. We are interested in $\mchi \gtrsim 300\gev$ to explain the AMS data, and shall fix $\Delta= 10^{-3}\mchi$. We shall vary other parameters in the ranges $10^{-4} \leq \mrho/\mchi \leq 10^{-1}$ and $10^{-2} \leq \alpha \leq 10^{-1}$. We discuss more about parameter values later.} The decay of the scalars is determined by $c_e,\,c_\mu,$ and $\Lambda$. We take the value of the SM vev as $v_H=256\gev$. 

\section{Cosmic History}\label{sec:history}
\begin{figure}[t]
\begin{center}
\includegraphics[width=0.95\columnwidth]{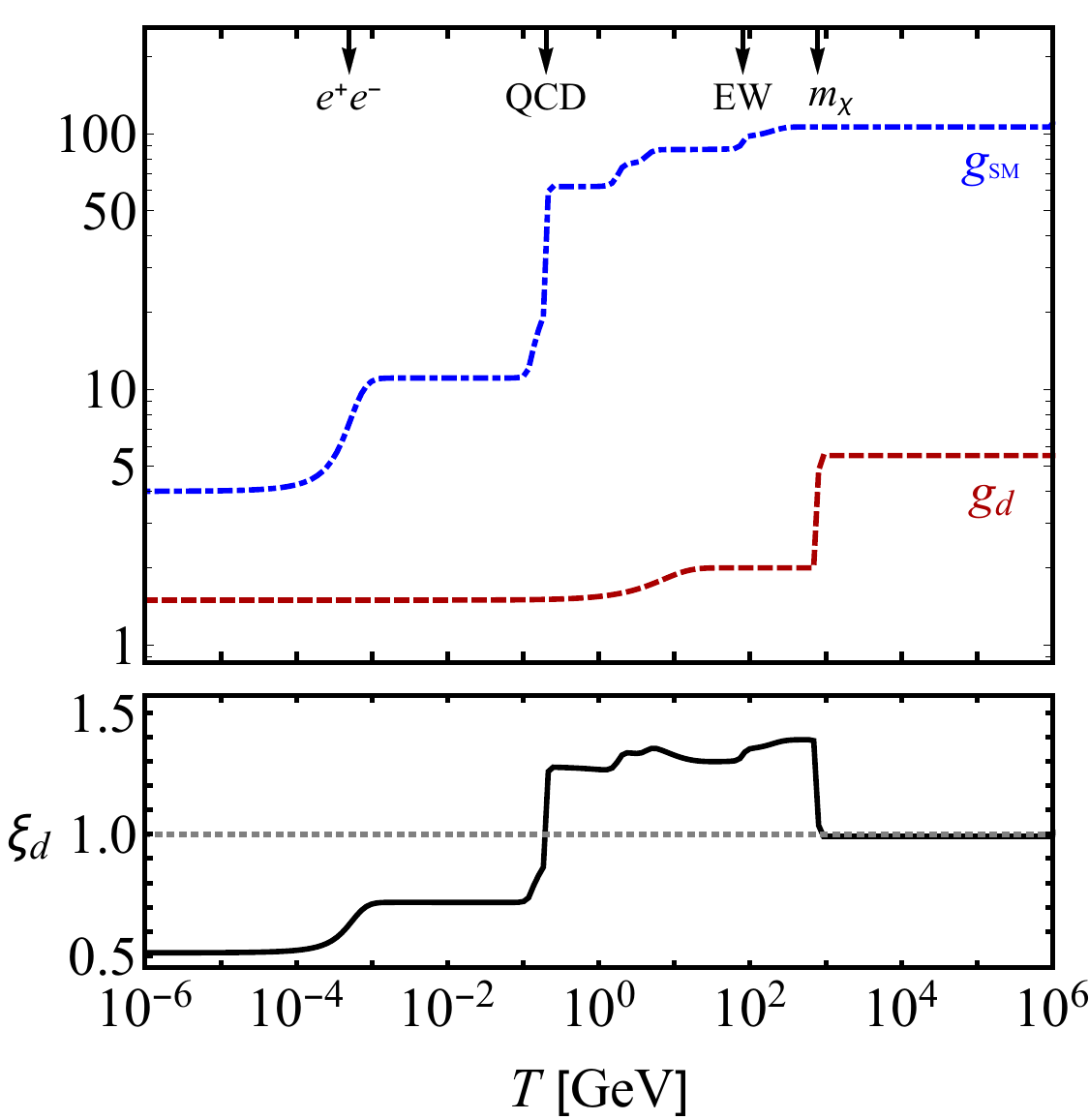}
\caption{Variation of the relativistic degrees of freedom in the SM $g_\text{SM}$ (dot-dashed blue) and the dark sector $g_d$ (dashed red). The lower panel shows the ratio of the dark sector temperature to the SM temperature $\xi_d$ (solid black).}\label{fig:xi}
\end{center}
\end{figure}

After the initial production of the dark sector particles through some mechanism like reheating, they decouple from the SM sector at a temperature $T_*$. We assume this to occur at a scale much higher relative to other energy scales in the theory. After decoupling, the SM and the dark sectors evolve independently with separate temperatures $T$ (SM) and $T_d$ (dark sector) with a ratio $\xi_d\equiv T_d/T$.  After the dark sector symmetry breaking at $T_d\simeq v_\Phi$, the dark sector consists of two Majorana fermions $\chi_1, \chi_2$ and two scalars $\rho, \eta$. The comoving entropy of this sector changes only during the decays of $\chi_2$ and $\rho$. In general $\xi_d$ can be written as 
\begin{equation}
	\xi_d(T)=\left(\frac{g_d(T_*)g_\text{SM}(T)}{g_d(T_d)g_\text{SM}(T_*)}\right)^{1/3}.
\end{equation}
Here $g_d(T)$ and $g_\text{SM}(T)$ are the relativistic degrees of freedom in the dark and the SM sectors, respectively, at temperature $T$. From Fig.\,\ref{fig:xi}, it is clear that the ratio $\xi_d$ is never too far away from unity. After $\rho$ and $\eta$ go out of chemical equilibrium, they decay. Note that $\rho$ could decay into $\eta$ if $\meta<\mrho/2$. However, as we will show later, this not the case in the region of the parameter space we are interested in. 
If the dark sector decouples from the visible sector early enough then $\rho,\,\eta$ do not affect the predictions for $N_\text{eff}$ and the Hubble parameter much during BBN, as was shown in \cite{Chu:2014lja}. Dark sector temperature $T_d$ rises after the DM decouples from the thermal bath. Later during the QCD phase transition at $T \sim 200\,{\rm MeV}$, a large amount of entropy is dumped into the SM thermal bath, heating it up.
This history depends on the relative values of $M$ and $T_*$ (see Fig.\,3 in Ref.\,\cite{Chu:2014lja}). In this paper, we shall only consider the case that the DM mass is above the QCD condensation scale.

We shall always assume that $M<v_\Phi$, so that the symmetry breaking occurs earlier than the chemical freeze-out of DM. The ensuing phenomenology was discussed in Ref.\cite{Weinberg:2013kea, Garcia-Cely:2013nin,Chu:2014lja}. \ad{Before thermal freeze-out, both $\chi_1$ and $\chi_2$ are present in the thermal bath. Afterwards, out-of-equilibrium $\chi_2$ particles decay and $\chi_1$ forms the DM abundance.} During relic annihilation, the interactions in Eq.(\ref{eq:int}) provide both annihilation and coannihilation, 
\begin{equation}\label{eq:2body}
 \begin{array}{ll}
  \chi_1\chi_1,\,\chi_2\chi_2 \to \rho\rho,\,\eta\eta & \quad\{\mathrm{annihilation}\}\,,\\[1ex] 
  \chi_1\chi_2\to \rho\eta & \quad\{\mathrm{coannihilation}\}\,.
 \end{array}
\end{equation}
Even though in this case the DM is lighter than the symmetry breaking scale $v_\Phi$, the scalar particle can be much lighter than $\chi_{1,2}$ if the hierarchy $v_\Phi>M>\mu_\Phi$ is maintained. 
\begin{figure}[t]
 \begin{center}
  \includegraphics[width=0.95\columnwidth]{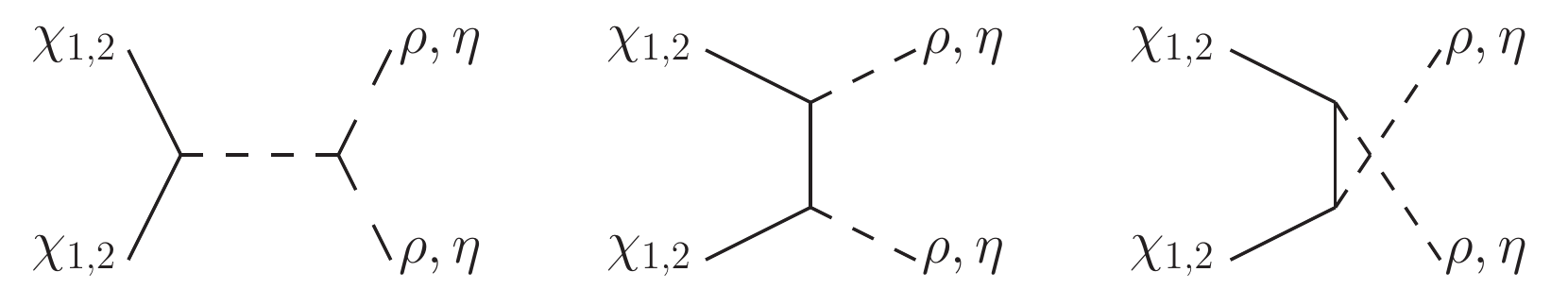}\\[4ex]
  \includegraphics[width=0.6\columnwidth]{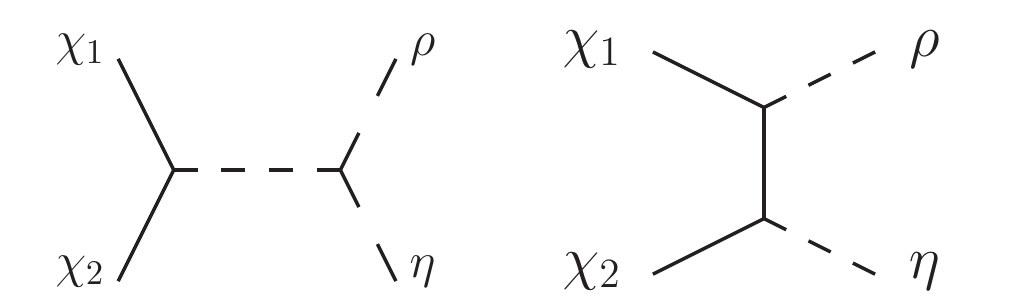}
  \caption{The Feynman graphs for DM annihilation and coannihilation.}\label{fig:feynamn_graph}
 \end{center}
\end{figure}

As $\chi_1,\chi_2$ are both Majorana particles, their annihilation into two scalars does not have any $s$-wave ($\ell=0$) component, and hence is $p$-wave ($\ell=1$) suppressed. But coannihilation is not subject to such suppression. 
Even after the DM is not in chemical equilibrium with the other particles in the dark sector, it exchanges kinetic energy with them as kinetic decoupling happens later than the chemical decoupling. In this case, the elastic scattering between $\chi_1$ and $\eta$ helps keep the DM in kinetic equilibrium with $\eta$. It was shown in Ref.\cite{Chu:2014lja} that a small mass split between the two DM states enhances this scattering cross section through a resonance. This delays the DM kinetic decoupling and may ameliorate the missing satellite problem\,\cite{Garrison-Kimmel:2013eoa,Bringmann:2006mu}. 


\section{Enhanced Annihilation}\label{sec:annihilation}
When the annihilating DM particles interact with each other through a long-range potential $V(r)$, then their initial plane-waves are modified. The Sommerfeld effect is the change in the annihilation rate resulting from this wavefunction modification. It is a nonperturbative effect. It is quantified by the Sommerfeld factor $S_\ell$ and is written as,
 \begin{equation}
  \sigma_\ell=S_\ell\,\sigma_{0\ell}\,.
 \end{equation}
 Here $\sigma_\ell$ and $\sigma_{0\ell}$ are the total and the perturbative cross-sections, respectively, for the orbital angular momentum $\ell=0,1,2,$ and so on. The process is enhanced if $S_\ell>1$, suppressed if $S_\ell<1$.

This effect is most pronounced when the DM particles are nonrelativistic. In this limit, it is useful to use the formalism of nonrelativistic effective theory (NREFT).
In NREFT, the wavefunction $\Psil$ of the heavy particles obey the \sch{} equation with the interaction potential $V(r)$. Relative to the range of $V(r)$, the actual annihilation process is short-range and is usually taken to be happening at the origin. Hence the Sommerfeld factor $S_\ell$ is defined as the change in the amplitude of the wavefunction $\Psil(\mathbf{r})$ at the origin due to the potential,
\begin{equation}
 S_\ell \equiv \left|\frac{\Psil(0)}{\Psil(0)_\text{free}}\right|^2\,.
\end{equation}
Here, $\Psil(\mathbf{r})_\text{free}$ is the wavefunction in the absence of any potential, \ie, the free particle plane wave. Depending on the nature of the mediator and the charges of the interacting particles, $V(r)$ can be either attractive or repulsive, causing enhancement or suppression, respectively, to the annihilation rate.

The possible two-particle states in the present case are the following.
\begin{equation}
\left .
  \begin{aligned}
    &|\chi_1\chi_1\rangle  \\[1ex]
    &|\chi_2\chi_2\rangle
  \end{aligned} \quad \right \} \quad \text{Annihilation}\nonumber
\end{equation}\vspace*{-0.cm}
 \begin{equation}
 \left .\quad
 \begin{aligned}
  &|\chi_1\chi_2\rangle \\[1ex]
  &|\chi_2\chi_1\rangle
 \end{aligned} \quad \right  \} \quad \text{Coannihilation}
 \end{equation}
The $\eta$-interaction leads to mixing between the two states in the annihilation space. But the annihilation and coannihilation subspaces remain decouple as neither $\rho$ nor $\eta$ interaction can couple one to the other. As mentioned before, the annihilation subspace does not have any $s$-wave process, but coannihilation has both $s$ and $p$-wave contributions. Although the two states for coannihilation consist of the same particles, we write them separately because it is easier to understand the transition from one to the other due to the $\eta$-exchange. We shall first compute the perturbative cross sections for these three processes using NREFT, and then calculate the Sommerfeld factors by solving the matrix \sch{} equations numerically.

\subsection{Annihilation matrices}
DM annihilation typically happens at a length scale $1/\mchi$. Therefore typical momentum exchange in such a process is $\sim \mathcal{O}(\mchi)$. In NREFT, the annihilation process cannot be described using tree level graphs. However, like in any EFT, all information regarding these `high energy' processes is contained the series of higher dimensional operators,
\begin{equation}\label{eq:lageff}
 \mathcal{L}_\mathrm{eff} = \sum_{i,j,k,l} c_{ij,kl}(\mchi)\,\overline{\chi}_i\chi_j\overline{\chi}_k\chi_l + \ldots\,.
\end{equation}
Here $c_{ij,kl}(\mchi)$ are the Wilson coefficients computed at the scale $\mchi$, and $\overline{\chi}_i\chi_j\overline{\chi}_k\chi_l$ are four-Fermi operators -- $\overline{\chi}_1\chi_1\overline{\chi}_1\chi_1$, $\overline{\chi}_1\chi_2\overline{\chi}_1\chi_1$, $\overline{\chi}_2\chi_2\overline{\chi}_1\chi_1$, and so on. The $c_{ij,kl}$ coefficients can be calculated by matching a four-point amplitudes in the full theory with the corresponding four-Fermi operator in the effective Lagrangian. 

To classify the effective operators according to the spin and angular momentum of the two-body states, we write the Dirac spinors of $\chi_i$ using the Pauli two-component spinors $\xi_i,\,\eta_i$ as
\begin{equation}
\begin{array}{rcl}
 	u_i(\pvec) &=& \sqrt{\dfrac{E_i+m_i}{2E}}\begin{pmatrix}
	\xi_i \\[1ex] \dfrac{\sigmavec\cdot\pvec}{E_i+m_i} \xi_i
	\end{pmatrix}\,,\\[5ex]
	v_i(-\pvec) &=& \sqrt{\dfrac{E_i+m_i}{2E}}\begin{pmatrix}
	\dfrac{-\sigmavec\cdot\pvec}{E_i+m_i} \eta_i \\[2ex] \eta_i
	\end{pmatrix}\,.\vspace{0.3cm}
\end{array}
\end{equation}
With these expressions for the spinors, we expand the operators in Eq.(\ref{eq:lageff}) in powers of $|\pvec|/\mchi$ to find the effective operators for each spin and angular momentum
\begin{equation}
 \mathcal{L}_\text{eff} = \sum_{i,j,k,l}\sum_{\ell,s} c_{ij,kl}(\spec)\,\mathcal{O}_{ij,kl}(\spec)\,,
\end{equation}
where $\mathcal{O}_{ij,kl}(\spec)$ are the effective operators consisting of the spinors $\xi_i,\eta_i$ corresponding to spin $s$ and angular momentum $\ell$ of the associated two-body states\,\cite{Bodwin:1994jh} (see Appendix\,\ref{sec:appendix_nr} for details about these operators).

Once the Wilson coefficients are known, the annihilation cross sections into the light scalars can be found by using the Cutkosky theorem: the annihilation cross section of a process $\chi_i\chi_j \to X_AX_B$ is proportional to the imaginary part of the Wilson coefficient of the operator $\chi_i\chi_j \to X_AX_B \to \chi_i\chi_j$\,\cite{Peskin:1995ev}. If $(\sigma v)_{\chi_i\chi_j \to X_AX_B} \equiv \Gamma\left(\chi_i\chi_j \to X_AX_B\right)$ is the annihilation rate then
\begin{equation}
 (\sigma v)_{\chi_i\chi_j \to X_AX_B} = 2\,\mathrm{Im}\left[c_{ij,ij}(\spec)\right]\,.
\end{equation}
In addition to $|\pvec|/\mchi$, we also expand the operators in powers of other dimensionless parameters, such as $\mrho/\mchi,\,\meta/\mchi$, and $\Delta/\mchi$. We computed the Wilson coefficients for both annihilation and coannihilation using the {\sc \mbox{FeynCalc}} package in {\sc Mathematica}\footnote{{\sc Mathematica} notebooks are available at this \href{https://github.com/anirbandas89/NREFT_Matching}{https URL}.} and classified them according to their spin and angular momentum\,\cite{Shtabovenko:2016sxi}. The procedure is described in detail in Appendix\,\ref{sec:appendix_nr}.

To the leading order in the dimensionless parameters mentioned above, we get the following annihilation matrices.
\begin{equation}\label{eq:ann_matrix}
 \begin{array}{l}
  \Gamma^\text{ann}_{\ell=1,\,s=1} = \dfrac{2\pi\alpha^2v^2}{\mchi^2}\begin{pmatrix}
                            +1&+1\\+1&+1
                           \end{pmatrix}\,,\\[4ex]
  \Gamma^\text{co-ann}_{\ell=0,\,s=1} = \dfrac{\pi\alpha^2}{16\mchi^2}\begin{pmatrix}
                            +1&-1\\-1&+1
                           \end{pmatrix}\,,\\[4ex]
  \Gamma^\text{co-ann}_{\ell=1,\,s=1} = \dfrac{\pi\alpha^2\mrho^2v^2}{16\mchi^4\Delta^2}\begin{pmatrix}
                            +1&+1\\+1&+1
                           \end{pmatrix}\,.
 \end{array}
\end{equation}
The $\ell=0$ coannihilation matrix has opposite signs for the off-diagonal entries relative to the diagonal elements. However, all elements of the $\ell=1$ matrices have similar sign. This fact is related to the particle exchange symmetry, and will be discussed in more detail when we describe our results. 

\subsection{Sommerfeld factors}
The nonrelativistic wavefunction of the DM particles obey the matrix \sch{} equation,
 \begin{equation}
 \label{eq:sch}
 \left[\mathcal{D}_{ab} + V_{ab}(r)\right]\left(\ul(r)\right)_{bc}=0\,.
 \end{equation}
 Here $\ul(r)$ is a $2\times 2$ matrix proportional to the radial parts $R(r) \left(=u(r)/r\right)$ of the full wavefunction $\Psil(\mathbf{r})$, and \mbox{$\mathcal{D}_{ab} = \left(-\frac{1}{2\mu}\frac{d^2}{dr^2} + \frac{\ell(\ell+1)}{2\mu r^2} - E\right)\delta_{ab}$} is the differential operator. The elements of $\ul(r)$ are the transition amplitudes $\left(\ul(r)\right)_{ab}\equiv \langle b|a\rangle$ of going from state $|a\rangle$ to $|b\rangle$, with $|a\rangle,|b\rangle$ being one of the three possible two-body states in Eq.(\ref{eq:2body}). As the particles are in scattering state before annihilation, the total energy of the $a^{\rm th}$ two-body state is
 \begin{equation}
  E_a \equiv \frac{1}{2}\mu_a v^2_\text{rel} = 2\mu_a v^2 = \frac{k_a^2}{2\mu_a}\,,
 \end{equation}
 where $v$ is the velocity of individual particles in the center-of-mass frame, and $k_a = \mu_av_\text{rel}$ is the momentum of the $a^\mathrm{th}$ two-body state. The potential $V(r)$ arises from the exchange of the scalars. It is given by
 \begin{equation}
  \begin{array}{ll}
   V(r) = \begin{pmatrix}
           V_{11} & V_{12}\\
           V_{21} & V_{22}
          \end{pmatrix}\,,
  \end{array}
 \end{equation}
 with $V_{11} = -\alpha\,e^{-\mrho r}/r$, $V_{12} = V_{21} = -\alpha\,e^{-\meta r}/r$, and $V_{22} = -\alpha\,e^{-\mrho r}/r + 2\Delta$ for annihilation and $V_{22} = -\alpha\,e^{-\mrho r}/r$ for coannihilation. The extra term $2\Delta$ in $V_{22}$ for annihilation comes from the mass gap between the two two-body states $|\chi_1\chi_1\rangle$ and $|\chi_2\chi_2\rangle$. No such mass gap exists for coannihilation.
 
 We follow Ref.\,\cite{Slatyer:2009vg} to compute the Sommerfeld factors for the multilevel DM. We quote the final expression for the Sommerfeld factor, $S_a\left(\spec\right)$, that is given as
 \begin{equation}\label{eq:sommerfeld}
 \left[\frac{(2\ell-1)!!}{k_a^\ell}\right]^2 \frac{(\tdag)_{ab}\mathcal{O}_{bc}(\spec)T_{ca}}{\mathcal{O}_{aa}(\spec)}\,.
 \end{equation}
No summation is implied over the repeated index $a$ in the denominator. The matrix $T$ is defined as the complex conjugate of the amplitude of the irregular solutions $v_\ell(r)$ of Eq.(\ref{eq:sch}) in the large $r$ limit,
 \begin{equation}
  \left(v_\ell(r\to\infty)\right)_{ab} = (\tdag)_{ab}\, e^{-ik_ar}\,.
 \end{equation}
 The detailed procedure to compute the $T$ matrix is given in the Appendix\,\ref{sec:appendix_sommerfeld}.

 \begin{figure}[t]
	\includegraphics[width=0.94\columnwidth]{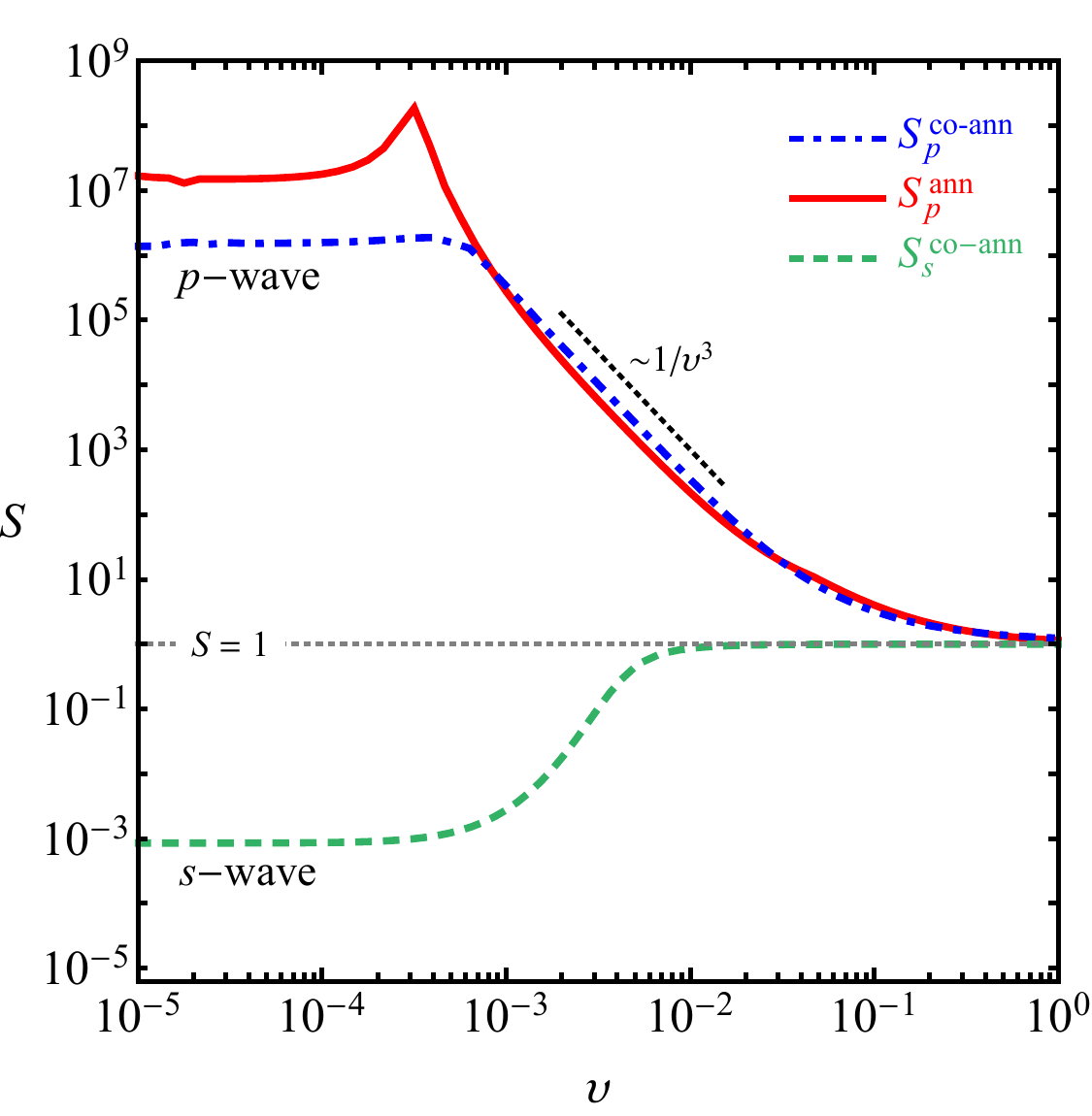}
	\caption{Sommerfeld factors for annihilation ($p$-wave) and coannihilation ($s$ and $p$-wave) as a function of DM velocity $v$.}\label{fig:sommerfeld_v}
\end{figure}

 We solved the \sch{} equation using two methods -- i) directly using {\tt NDSolve} in {\sc Mathematica}, and ii) using the \emph{variable phase method}\,\cite{Ershov:2011zz,Beneke:2014gja}. The direct method works well as long as the kinetic energy of the incoming particles is above the threshold for the $|\chi_2\chi_2\rangle$ \mbox{state, \ie,}
 \begin{equation}
  E \geq 2\Delta\,.
 \end{equation}
 When the incoming particles are below threshold, the wavefunctions for $|\chi_2\chi_2\rangle$ final state are not scattering state anymore. In this case, the direct method fails as it simultaneously tries to solve for scattering and bound state solutions within a single system. Ref.\,\cite{Beneke:2014gja} proposed that the variable phase method can be used in such cases. Instead of solving for the full solution, this method solves for the modification in the wavefunction from free particle solution because of the long range potentials. We have verified the results from the variable phase method with direct method solutions for above threshold parameters. We found excellent match between the results from two methods.

\begin{figure}[t]
	\includegraphics[width=0.96\columnwidth]{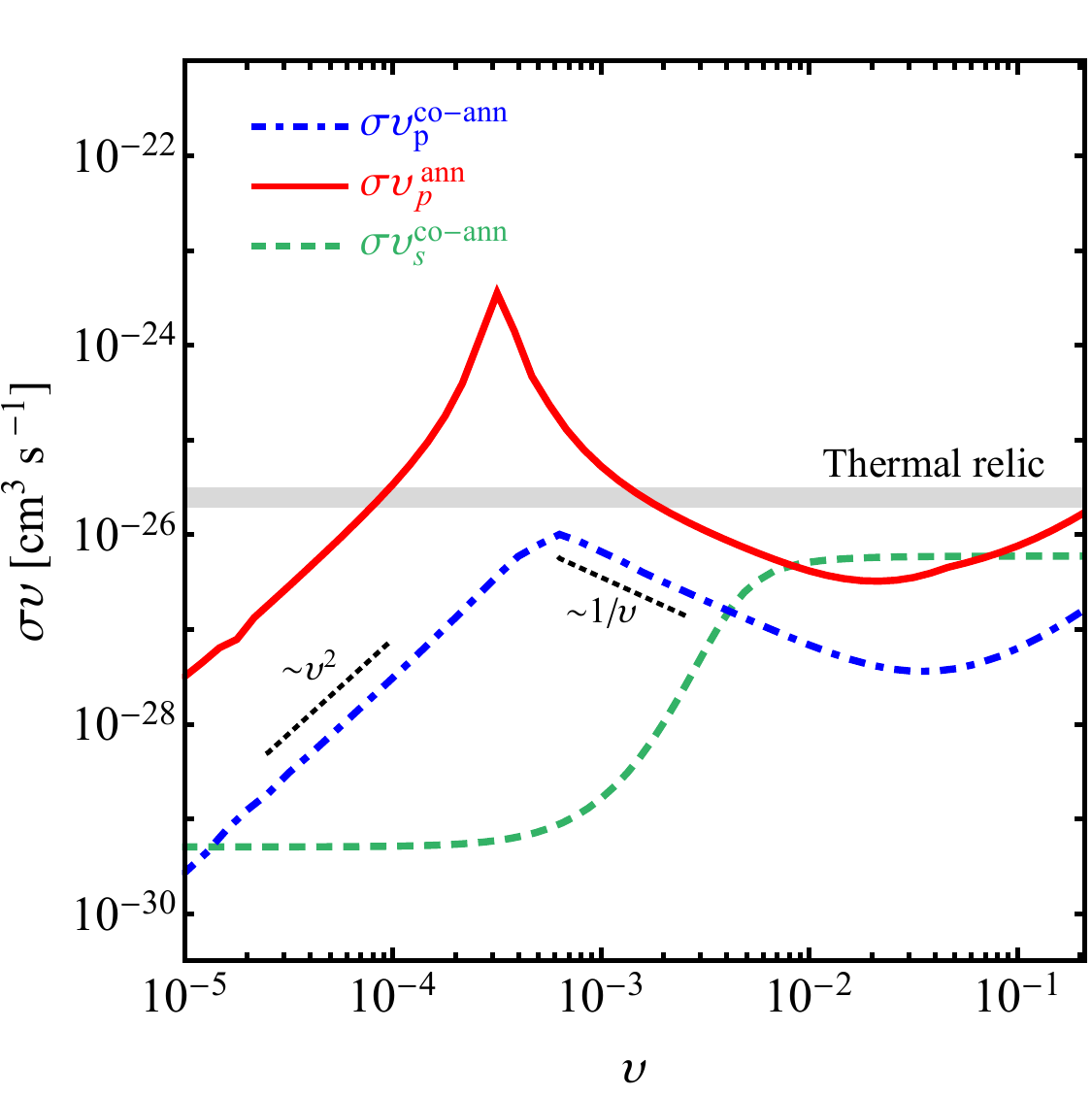}
	\caption{Sommerfeld enhanced $p$-wave annihilation rate $S_p(\sigma v)_p$ (solid red) as a function of DM velocity $v$. The $s$ and $p$-wave coannihilation rates are shown for comparison as the dashed green and dot-dashed blue curves. The thermal relic annihilation rate $\sigma v_\mathrm{relic}=(2-3)\times 10^{-26}\cms$ is shown as a gray band.}\label{fig:rate_v}
\end{figure}

To reduce the number of parameters, we fix $\meta = 0.7\mrho$ for all the results shown in this paper. \ad{A different choice for $\meta$ does not change our results qualitatively.} In Fig.\,\ref{fig:sommerfeld_v} and \ref{fig:rate_v}, we show the Sommerfeld factors and the annihilation rates, respectively, as functions of DM velocity $v$. For the figures in the next two subsections, we choose a representative set of values for the parameters: $\alpha = 0.0373,\,\mrho/\mchi = 0.0023,\,\Delta/\mchi = 0.001,\,v=100\,\mathrm{km\,s^{-1}}$ which approximately corresponds to the point marked with orange asterisk in Fig.\,\ref{fig:contour}. 

Both $p$-wave annihilation and coannihilation show large enhancement in the small velocity limit in Fig.\,\ref{fig:sommerfeld_v}. The enhancement factor has a $1/v^3$-dependence in the intermediate velocity regime. The difference between the annihilation and coannihilation factors is due to the mass gap between the two states. However, the $s$-wave Sommerfeld factor is less than one. The $p$-wave annihilation and the $s$-wave coannihilation rates are shown in Fig.\,\ref{fig:rate_v}. The $\sim v^2$-scaling of the bare $p$-wave cross section yields the $\sim v^2$ and $\sim v^{-1}$ behavior of $\sigma v_p$ in the small and intermediate velocity regimes, respectively. We note that the relic annihilation cross section is dominated by annihilation because of a residual Sommerfeld enhancement factor $S_p\simeq 2$ even for $v_\mathrm{relic}\simeq 0.2$. Therefore, $p$-wave annihilation process contributes more by a factor of $\sim 2$ than coannihilation in the early Universe. This more carefully computed result disagrees somewhat with the conclusions made in previous papers\,\cite{Garcia-Cely:2013nin,Chu:2014lja,Das:2016ced}. However, we must note that at this large velocity $v\sim 0.2$, the partial wave expansion of the cross section may not be very accurate as higher partial wave contributions become important. Fig.\,\ref{fig:sommerfeld_mrho} shows the $\mrho$-dependence of the factors. In addition to a large overall enhancement for $p$-wave, a resonance feature is also present for certain values of $\mrho$. We shall explain the origin of these features in the rest of this section.

\subsection{Particle exchange symmetry}
We shall use the particle exchange symmetry following the Ref.\,\cite{Das:2016ced}. Suppose, $A$ and $B$ are two fermions. They can form two two-body states, namely $|AB\rangle$ and $|BA\rangle$ of total angular momentum $\ell$ and spin $s$. However, these two states consist of the same set of particles and are related to each other through
\begin{equation}\label{eq:exchange}
 |AB\rangle = (-1)^{\ell+s} |BA\rangle\,.
\end{equation}
This factor has three components: $(-1)^\ell$ from relative angular momentum, $(-1)^{s+1}$ from their spins, and $(-1)$ due to Wick exchange of two fermions. 
\begin{figure}[t]
 \includegraphics[width=0.93\columnwidth]{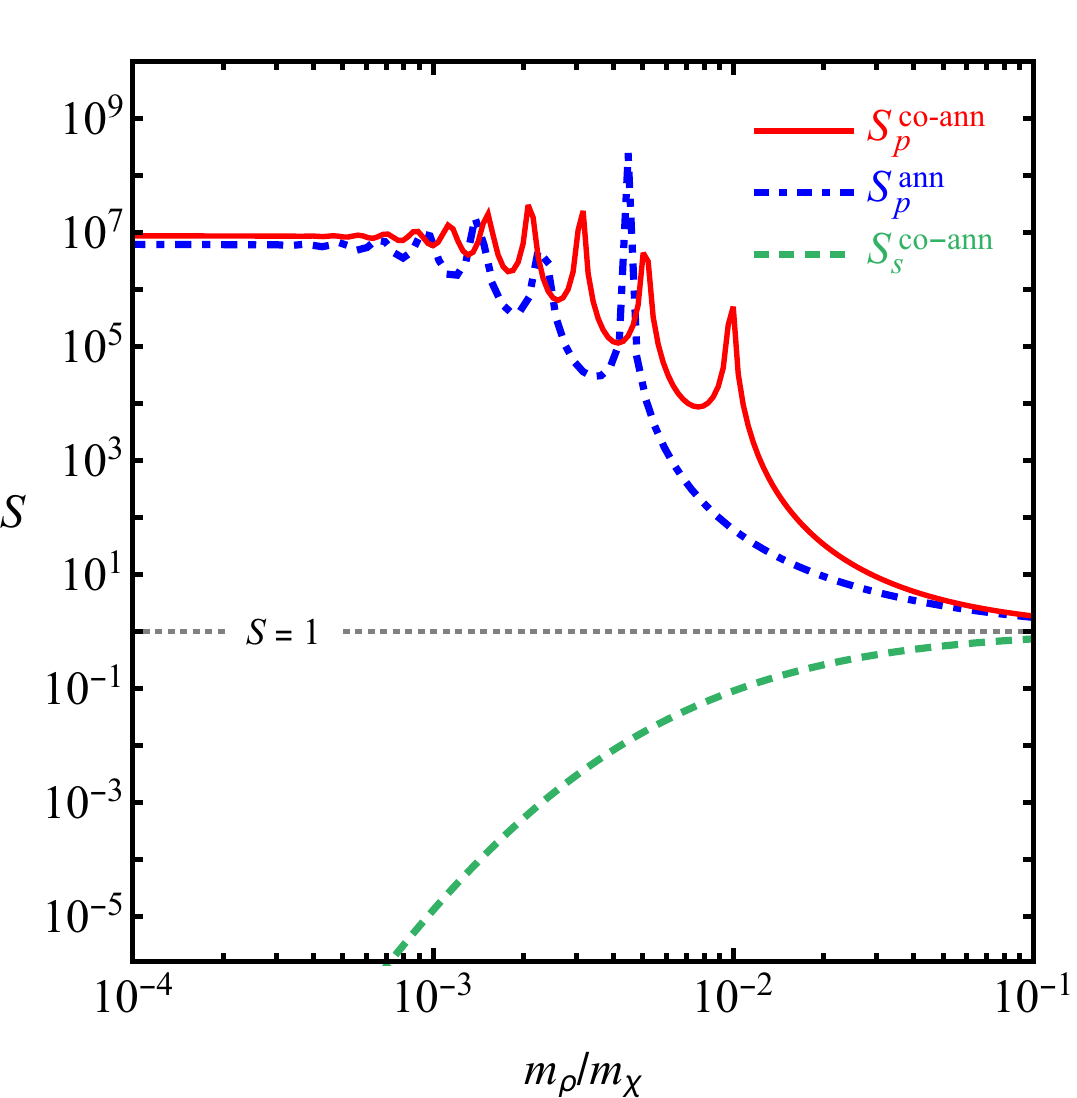}
 \caption{Sommerfeld factors for annihilation ($p$-wave) and coannihilation ($s$ and $p$-wave) as a function of the diagonal mediator mass $\mrho$.}\label{fig:sommerfeld_mrho}
\end{figure}

\begin{figure}[t]
 \begin{center}
  \includegraphics[width=0.95\columnwidth]{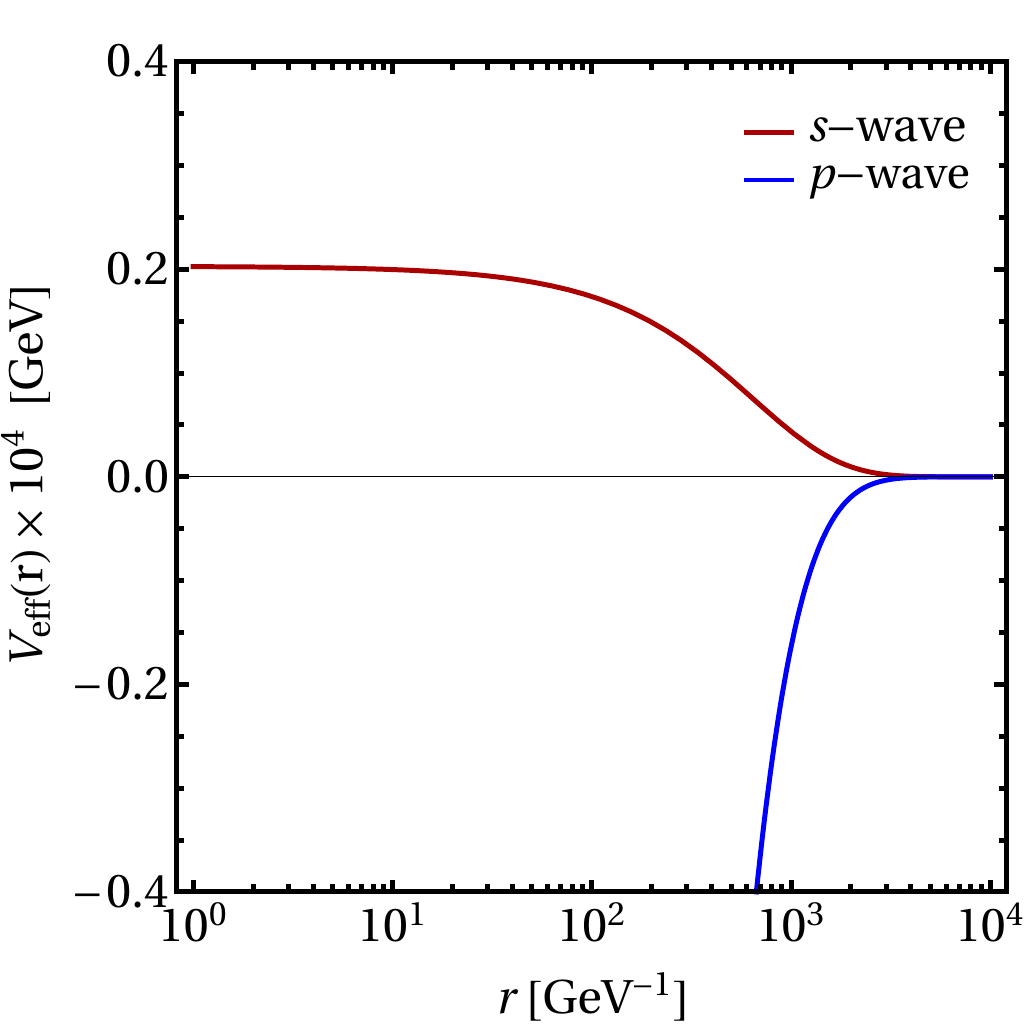}
  \caption{The 1D effective potentials for coannihilation in Eq.(\ref{eq:effective}).}\label{fig:effective}
 \end{center}
\end{figure}

We can directly apply this to the coannihilation channel in the present model implying that two states $|\chi_1\chi_2\rangle$ and $|\chi_2\chi_1\rangle$ in the matrix \sch{} equation are related to each other,
\begin{equation}
 |\chi_1\chi_2\rangle = (-1)^{\ell+s} |\chi_2\chi_1\rangle\,.
\end{equation}
It further implies that we can combine the two equations into a single equation with an effective potential which is a linear combination of the diagonal and off-diagonal potentials (see Appendix\,\ref{sec:appendix_onelevel} for details),
\begin{equation}
 \veff = V_{11} + (-1)^{\ell+s} V_{12}\,.
\end{equation}
Therefore $\veff$ has the following forms for the two cases,
\begin{equation}\label{eq:effective}
 \begin{array}{l}
  \veff^{\ell=0,\,s=1} = -\dfrac{\alpha\,e^{-\mrho r}}{r} + \dfrac{\alpha\,e^{-\meta r}}{r}\,,\\[2ex]
  \veff^{\ell=1,\,s=1} = -\dfrac{\alpha\,e^{-\mrho r}}{r} - \dfrac{\alpha\,e^{-\meta r}}{r}\,.
 \end{array}
\end{equation}
We note that in the $\ell=0$ case, the difference between the two potentials is acting as the effective 1D potential. When $r \ll 1/\mrho$ and $1/\meta$, $\veff$ saturates at the value ($\mrho-\meta$). For larger $r$, it gradually decreases to zero never becoming negative. The nature of the net potential thus becomes \emph{repulsive} (See Fig.\,\ref{fig:effective}). However for $\ell=1$, the diagonal and off-diagonal potentials are added with the same sign, and hence yield an \emph{even stronger attractive} potential. This explains the behaviour of the different Sommerfeld factors in Figs.\,\ref{fig:sommerfeld_v} and \ref{fig:sommerfeld_mrho}. The effective repulsive nature of $\veff$ leads to the suppression in $S^\text{co-ann}_s$.

\subsection{Velocity dependence \& resonances}
The $1/v^3$-dependence of $S_p$ in Fig.\,\ref{fig:sommerfeld_v} can be understood by taking the limit $\mrho/\mchi \ll 1$ and $\alpha/v \gg 1$. In this Coulomb limit, an analytic expression for $S_\ell$ for general angular momentum $\ell$ is given by\,\cite{Cassel:2009wt}
\begin{equation}
 S_\ell = S_0 \prod_{j=1}^\ell \left(1 + \frac{\alpha^2}{j^2v^2}\right)\,,
\end{equation}
where $S_0$ is the $s$-wave factor in the Coulomb approximation
\begin{equation}
 S_0 = \frac{\pi\alpha/v}{1-e^{-\pi\alpha/v}}\,.
\end{equation}
Therefore for small velocity, $S_\ell \sim 1/v^{2\ell+1}$. However for finite $\mrho$, the Sommerfeld factor does not grow indefinitely for smaller velocity. Instead, it saturates to a constant value when the de Broglie wavelength of the DM particles gets much longer than the range of the potential, \ie, approximately when the condition $\mu_1v/\mrho \ll 1$ is met. All of these features are evident in Fig.\,\ref{fig:sommerfeld_v}.

Resonances from virtual bound states arise and cause large Sommerfeld enhancement as can be seen in Fig.\,\ref{fig:sommerfeld_mrho}. This occurs whenever the range of the potential matches a multiple $n$ of the Bohr radius $1/(\alpha\mchi)$ of the DM particles, \ie,\,\cite{Cassel:2009wt}
\begin{equation}
 \frac{\alpha\mchi}{\kappa\mrho} = n^2, \qquad n = 1,2,3,\ldots\,.
\end{equation}
Here $\kappa \simeq \pi^2/6$. Note that this resonance condition is only approximately true in the present model as we have two mediators with slightly different masses. Also the resonance positions are shifted by the mass gap $\Delta$\,\cite{Slatyer:2009vg}.

A consequence of the nonmonotonic velocity dependence of the $p$-wave Sommerfeld factor is that it predicts different annihilation rates of DM in astrophysical objects of different size. In the net $p$-wave annihilation rate, the perturbative cross section provides a velocity scaling $\sim v^2$. Therefore, $S_p(\sigma v)_p$ increases as $\sim v^2$ for small velocity, reaches a maximum, and then falls as $\sim 1/v$ as shown in Fig.\,\ref{fig:rate_v}. The position of the maximum  annihilation depends on the ratio $\mrho/\mchi$. 
Hence this model naturally predicts large DM annihilation signal in the galaxies, but very small signal from the dwarf galaxies.

\section{Phenomenological Consequences}
\label{sec:pheno}

\subsection{Galactic positron excess}
The AMS-02 and several other experiments have observed an excess in the positron spectrum from the galaxy, as mentioned in the introduction. DM annihilation in our model can be used to explain such observation. To this end, we show the variation of the $p$-wave annihilation rate $(\sigma v)_p$ in the Milky Way with $\mrho/\mchi$ and $\alpha$ in Fig.\,\ref{fig:contour}. The cross section is enhanced for large $\alpha$ and small $\mrho/\mchi$ ratio, with a resonance feature as discussed above. The points within the overlaid white band yield a relic annihilation cross section within $(2-3)\times 10^{-26}\cms$ to satisfy the relic abundance constraint\,\cite{Steigman:2012nb}. 

The points marked with asterisks yield DM annihilation cross section $\sigma v > 10^{-24}\cms$, sufficient to explain the AMS-02  positron flux excess, without running into any problems with thermal relic or dwarf galaxy constraints. The typical annihilation cross section in dwarf galaxies is $\lesssim 6\times 10^{-28}\cms$, which is well below the Fermi-LAT bound in the $e^+e^-$ and $\mu^+\mu^-$ channels\,\cite{Ackermann:2015zua}. Because of the suppression in the $s$-wave channel as shown in Fig.\,\ref{fig:rate_v}, DM annihilation rate during recombination era is also predicted to be much smaller than the current experimental bound\,\cite{Kawasaki:2015peu}. We show the positron flux from such a representative point in Fig.\,\ref{fig:positronflux1}. We used the publicly available code \textsc{PPPC4DMID} to compute the positron flux spectrum in the $\chi_1\chi_1\to VV\to 4e,\,4\mu$ channel, where the scalar $V$ represents $\rho$ and $\eta$, from the galaxy after diffusion\,\cite{Cirelli:2010xx}.

\begin{figure}[t]
	\begin{center}
		\includegraphics[width=0.95\columnwidth]{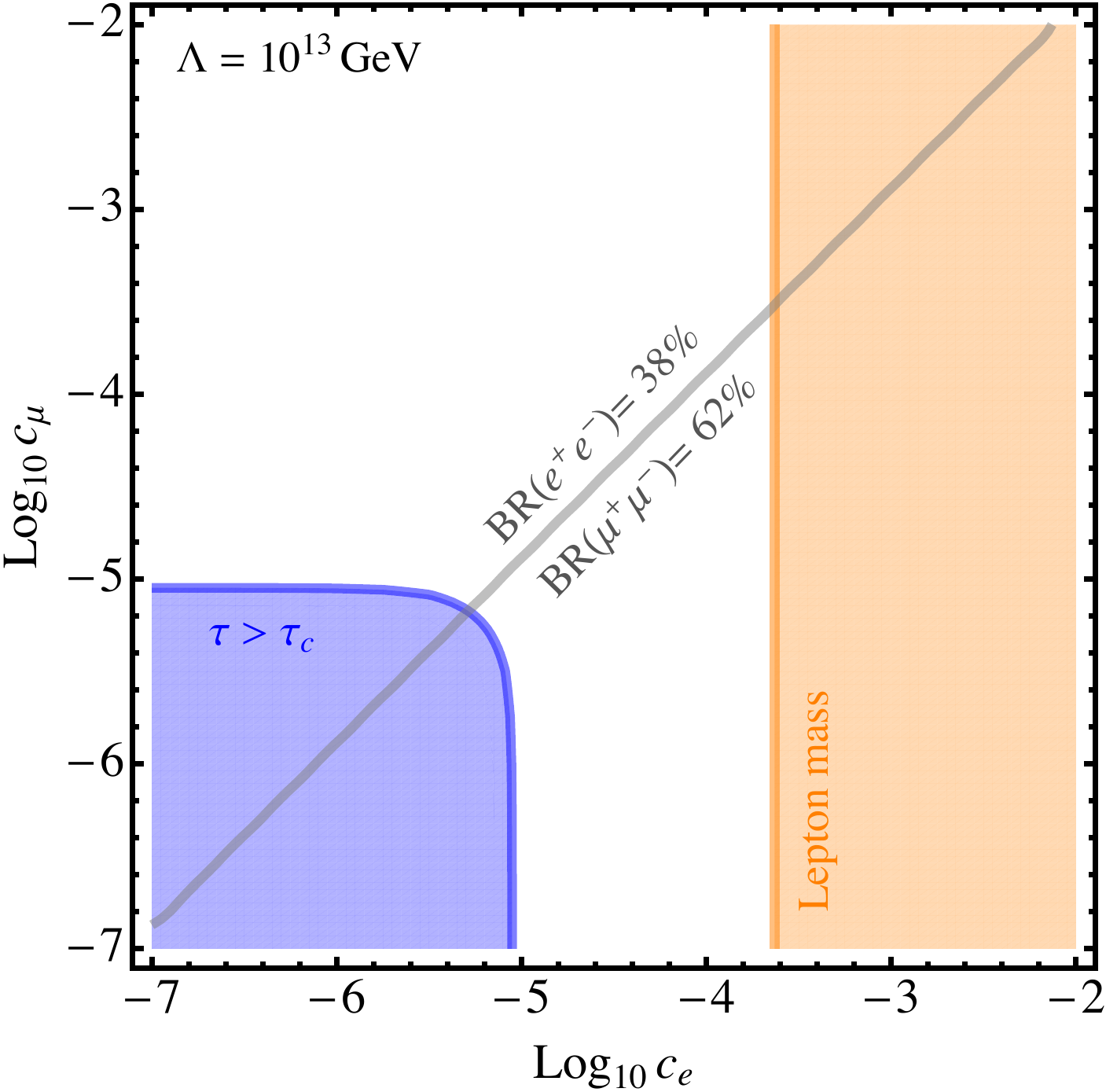}
		\caption{The model parameter space in the $c_e-c_\mu$ plane for $\Lambda=10^{13}\gev$. The gray line is the contour of $\mathrm{BR}(e^+e^-)=38\%$ and $\mathrm{BR}(\mu^+\mu^-)=62\%$. In the blue-shaded region, $\rho$ and $\eta$ decay times are larger than $\tau_c$. See text for definition. The orange-shaded region is excluded by lepton mass measurement.}
		\label{fig:ce_cmu1}
	\end{center}
\end{figure}
\begin{center}
 \begin{table}[b]
 \centering
 \caption{The dark sector and background parameters (see Eq.(\ref{eq:background})) used in Fig.\,\ref{fig:positronflux1}.}\label{tab:ams_param}
 \vspace{0.2cm}
 \begin{tabular}{M{2.3cm}M{5cm}}
 \hhline{--}
 Parameter & Value\\ \hline
 $\mchi$ & $780\gev$\\ 
 $\sigma v$ & $4.63\times 10^{-24}\cms$\\ 
 $\text{B.R.}(e^+e^-)$ & $38\%$\\
 $\text{B.R.}(\mu^+\mu^-)$ & $62\%$\\
 Halo profile & Einasto\\
 $C_d$ & $6.42\times 10^{-2}\,(\text{GeV}\,\text{m}^2\,\text{s}\,\text{sr})^{-1}$\\
 $\phi_{e^+}$ & $0.869\gev$\\
 $\gamma_d$ & $-3.6$\\
 $E_1$ & $7\gev$\\
 \hhline{--}
 \end{tabular}
\end{table}
\end{center}
An astrophysical background of positron flux was assumed, following Ref.\,\cite{Aguilar:2019owu},
\begin{equation}\label{eq:background}
 F_{e^+}(E) = C_d^2 \left(\frac{E^2}{\hat{E}^2}\right) \left(\frac{\hat{E}}{E_1}\right)^{\gamma_d}\,,
\end{equation}
where $\hat{E} = E + \phi_{e^+}$ is the positron energy in the interstellar space. 
The dark sector and background parameters used in Fig.\,\ref{fig:positronflux1} are listed in Table\,\ref{tab:ams_param}. For $\mchi=780\gev$, the required mediator mass is about $\mrho\simeq 1.5\gev$ (see Fig.\,\ref{fig:contour}). Therefore $\rho$ and $\eta$ can decay to electron and muon. Note that we are not using the background parameters quoted in Ref.\,\cite{Aguilar:2019owu} as this is an independent model. The fit shown in Fig.\,\ref{fig:positronflux1} corresponding to the parameter values quoted in Table\,\ref{tab:ams_param} were found by a crude scan over the parameter space. Although the fit of the theoretical curve to the data appears acceptable, the goodness of the fit is rather modest \ad{with $\chi^2/\mathrm{d.o.f} = 2.07$. }

\begin{figure}[t]
	\begin{center}
		\includegraphics[width=0.95\columnwidth]{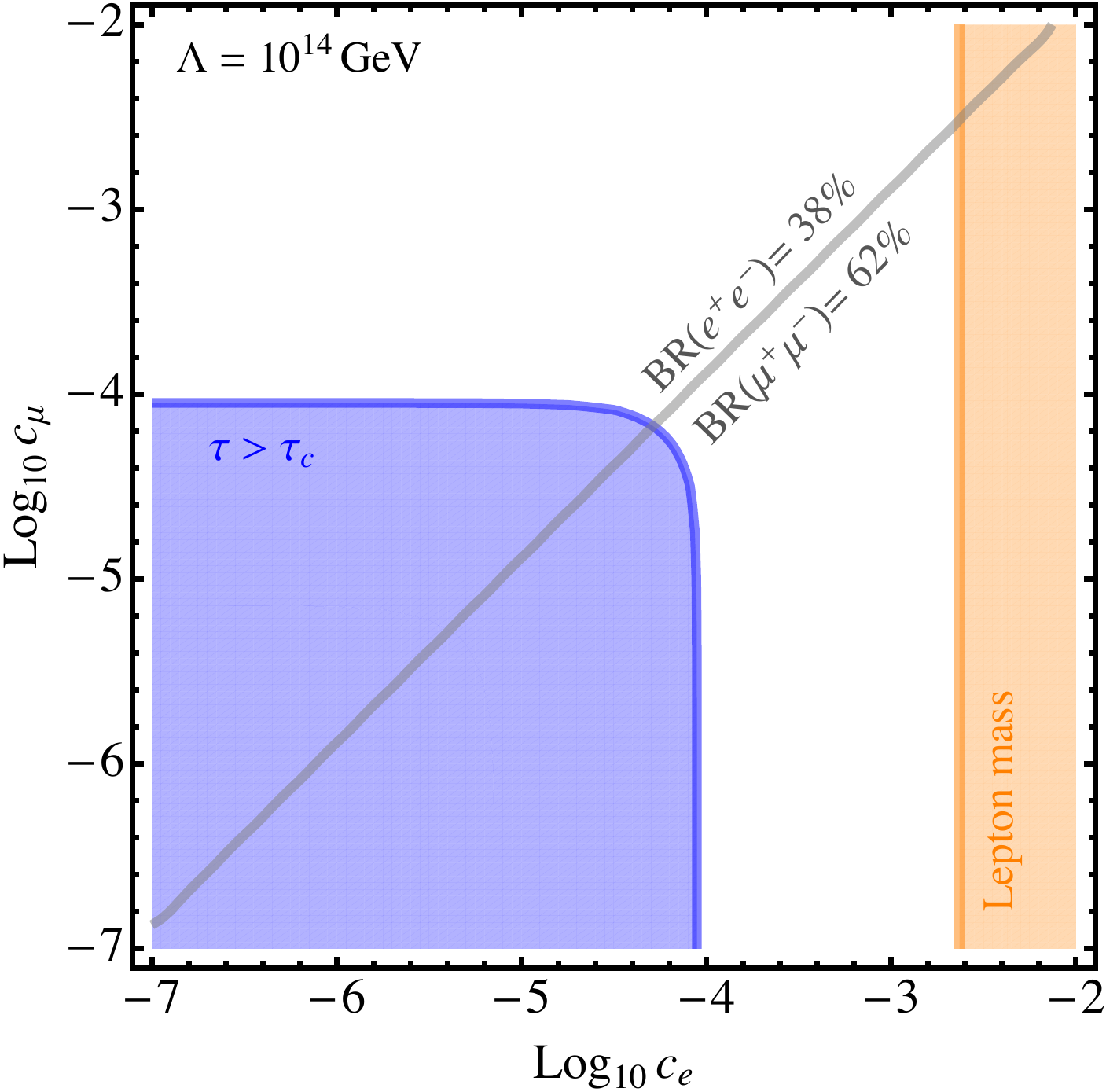}
		\caption{Same as Fig.\,\ref{fig:ce_cmu1} but with $\Lambda=10^{14}\gev$. The position of the gray line does not change as the branching ratio does not have any $\Lambda$-dependence.}
		\label{fig:ce-cmu2}
	\end{center}
\end{figure}

\ad{Three dark sector parameters $\mchi, \sigma v$, and BR$(e^+e^-)$ determine the shape and position of the bump in the positron flux. DM mass $\mchi$ fixes its position as the DM particles are highly nonrelativistic today, $\sigma v$ and BR$(e^+e^-)$ together determine the amplitude and the spread of the peak. It might seem from Fig.\,1 that increasing $\mchi$ could give a better fit to the last three data points in the high energy end. However, as the peak shifts towards higher energy, the fit in the intermediate energy becomes poor. Increasing $\sigma v$ has the effect of increasing the overall amplitude of the flux. The height of the peak is also partially controlled by BR$(e^+e^-)$, decreasing this parameter lowers the height increasing its width. We used the Einasto profile for the MW halo,
\begin{equation}
\mathrm{Einasto:}\quad \rho(r) = \rho_s \exp\left[-\dfrac{2}{\alpha}\left(\left(\dfrac{r}{r_s}\right)^\alpha-1\right)\right]\,.
\end{equation}
We used the default values for the parameters implemented in \textsc{PPPC4DMID}: $\rho_s=0.033\gev \mathrm{cm^{-3}}, r_s=28.44\kpc$, and $\alpha=0.17$\,\cite{Boudaud:2014qra}. With these parameters, the local DM density is $0.32\gev \mathrm{cm^{-3}}$.
In this context, we want to mention the effect of the morphological dependence of the $J$-factor on the positron flux\,\cite{Boddy:2018ike}. Averaging the annihilation rate over velocity will decrease the effective rate, given that our parameters correspond to a resonance at some $v$. However, several factors mitigate this effect. The Einasto density profile that we use is expected to give a smaller morphological dependence than a cuspy profile used in Ref.\,\cite{Boddy:2018ike}. Further, the AMS-02 data for the positron flux is not directional, rather it is the total flux integrated over all directions as observed by the spectrometer, thus diluting this effect. To arrive a more quantitative conclusion, however, the calculation of Ref.\,\cite{Boddy:2018ike} must be repeated for the nonmonotonic velocity dependence of the Sommerfeld factor.}

\ad{The $\chi^2/\mathrm{d.o.f}$ reported here can be easily improved. For example, fitting the low and intermediate energy data points more precisely, at the cost of fitting the higher energy data less well, improves the $\chi^2$ because of the relatively large error bars for the high energy data points. If more data shrinks these error bars (without changing the central values), they would easily rule out such parameters. Thus we have chosen to show a benchmark point that will survive such a scenario to a greater extent. We also note the glitches in the AMS-02 data around $E \sim 100\gev$ which worsens the fit. We believe more data in future will settle these issues.}

The annihilation products $\rho,\,\eta$ couple to the visible sector through the charged leptons $e$ and $\mu$ with branching ratios $38\%$ and $62\%$, respectively, to get an acceptable fit. One may ask if it is possible to get these branching ratios within our model. The $\Phi H\bar{l}_Ll_R$ term, shown in Eq.(\ref{eq:lag}), leads to decay of $\rho$ and $\eta$ generated in the DM annihilations. We compare their decay times with $\tau_c\,(\simeq 47\text{\,yrs})$ which is the typical time required by them to escape the MW without any scattering. The decay widths are parametrized by $c_e,c_\mu$, and $\Lambda$. For $c_e, c_\mu \gtrsim 10^{-5}$ and $\Lambda=10^{13}\gev$, their lifetimes $\tau_{\rho,\eta}$ are less than $\tau_c$, as shown in Fig.\,\ref{fig:ce_cmu1} and \ref{fig:ce-cmu2}. For our purpose,  their decay can be assumed to be prompt.

\ad{In passing, we want to emphasize that the aim of this work is to demonstrate the possibility of explaining the positron flux excess using $p$-wave Sommerfeld enhanced DM annihilation in the MW. Given the moderate goodness of fit, we do not claim that this particular DM model is an excellent candidate to explain the excess. Rather, we want to point out the uniqueness of this solution as it can accommodate the required large annihilation cross section without conflicting the dwarf galaxy or CMB data. We hope that this work will spawn a new direction in model-building and associated cosmic ray data analysis. Note that the \emph{selective Sommerfeld enhancement} described in this paper is a general feature of any multilevel self-interacting DM models with light mediator.}
\subsection{Other BSM effects}

The $\Phi H\bar{l}_Ll_R$ operator also modifies $e$ and $\mu$ masses, as well as their coupling with the SM Higgs $h$. The electron and muon masses are predicted to change by an amount $\sim c_lv_\Phi v_H/\Lambda$. The present experimental precisions of their mass measurement are very high, at the level of \mbox{$\sim 10^{-8}$}\,\cite{Mohr:2015ccw}. Specifically, electron mass measurement constrains our parameter space significantly. The exclusion regions due to this are shown as orange-shaded regions in Figs.\,\ref{fig:ce_cmu1} and \ref{fig:ce-cmu2}. The branching ratio of Higgs decay into muons is also affected. However, the resulting change in the Higgs signal strength in the $\mu^+\mu^-$ channel is very small relative to the current measurement, hence we do not consider it any further\,\cite{Aaboud:2017ojs,	Sirunyan:2018hbu}. 

In Figs.\,\ref{fig:ce_cmu1} and \ref{fig:ce-cmu2}, we show the contour of $\mathrm{BR}(e^+e^-)=38\%$ and $\mathrm{BR}(\mu^+\mu^-)=62\%$ with a gray solid line. Everywhere on the gray lines, the model produces a good fit to data, though the blue shaded region in the lower left is disfavored because the scalars decay too slowly ($\tau>\tau_c$) and the orange shaded region in the top right is disfavored because the fractional shift to lepton masses is larger than $10^{-8}$.

 \begin{figure}[t]
  \begin{center}
   \includegraphics[width=0.94\columnwidth]{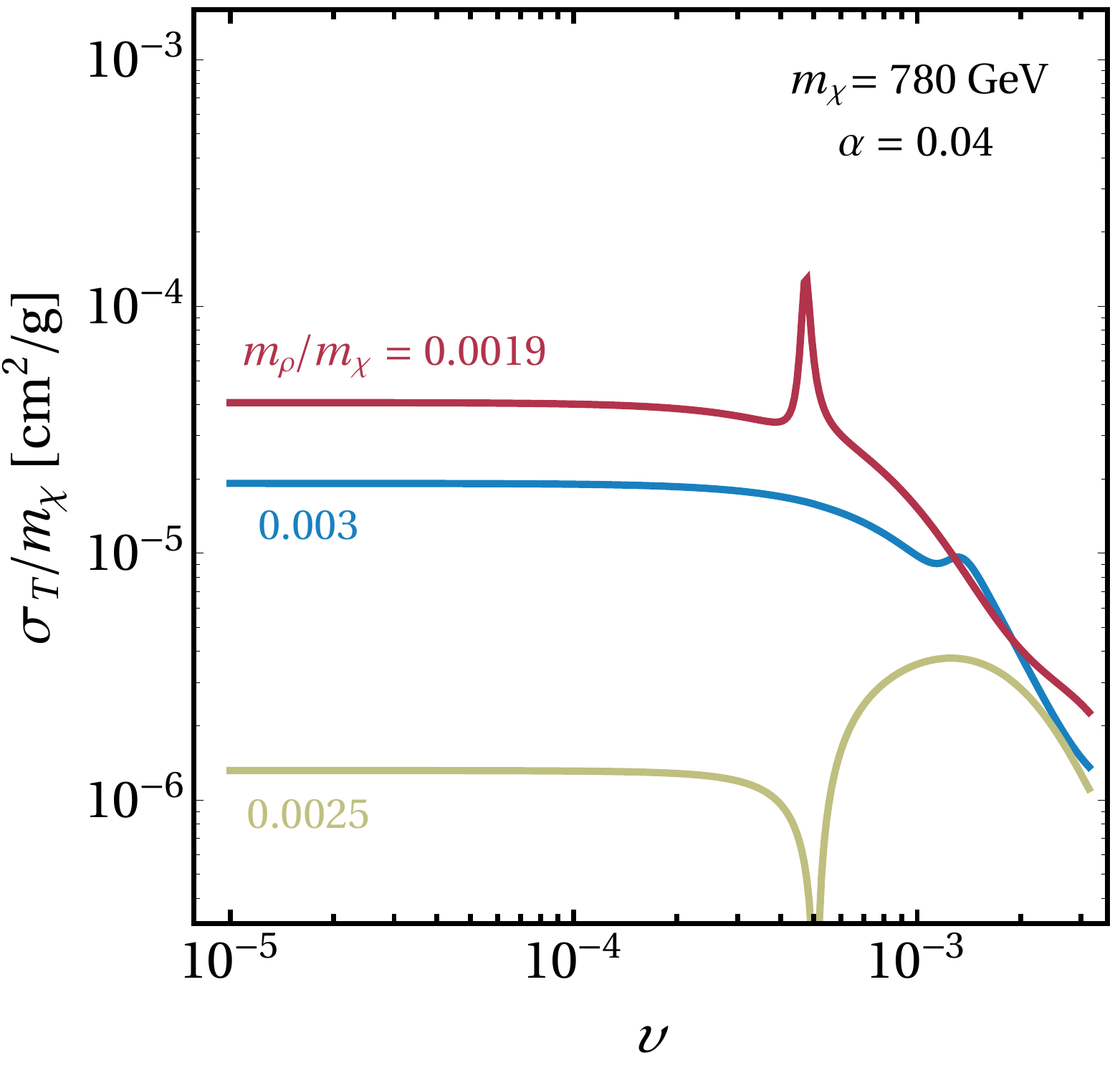}
   \caption{The elastic transfer cross section $\sigma_\text{T}$ for $\chi_1\chi_1\to\chi_1\chi_1$ as a function of DM velocity. The mass gap is taken to be small, $\Delta = 0.1\mev$ to ensure numerical stability. Larger $\Delta$ changes the cross section at most by a factor of a few.}\label{fig:scattering_v}
  \end{center}
 \end{figure}

\subsection{Self-scattering}\label{sec:scattering}
The long range attractive potential would enhance DM self-scattering cross section as well. Large self-scattering cross section, say $\sigma/\mchi \simeq 1-10\cmg$, may affect DM halo formation and produce core at the center\,\cite{Spergel:1999mh,Loeb:2010gj}. At the same time, the Bullet cluster observation puts an upper limit $\sigma/\mchi \lesssim 0.1\cmg$ on this cross section\,\cite{Clowe:2006eq}.
\begin{figure}[t]
 \begin{center}
  \includegraphics[width=0.95\columnwidth]{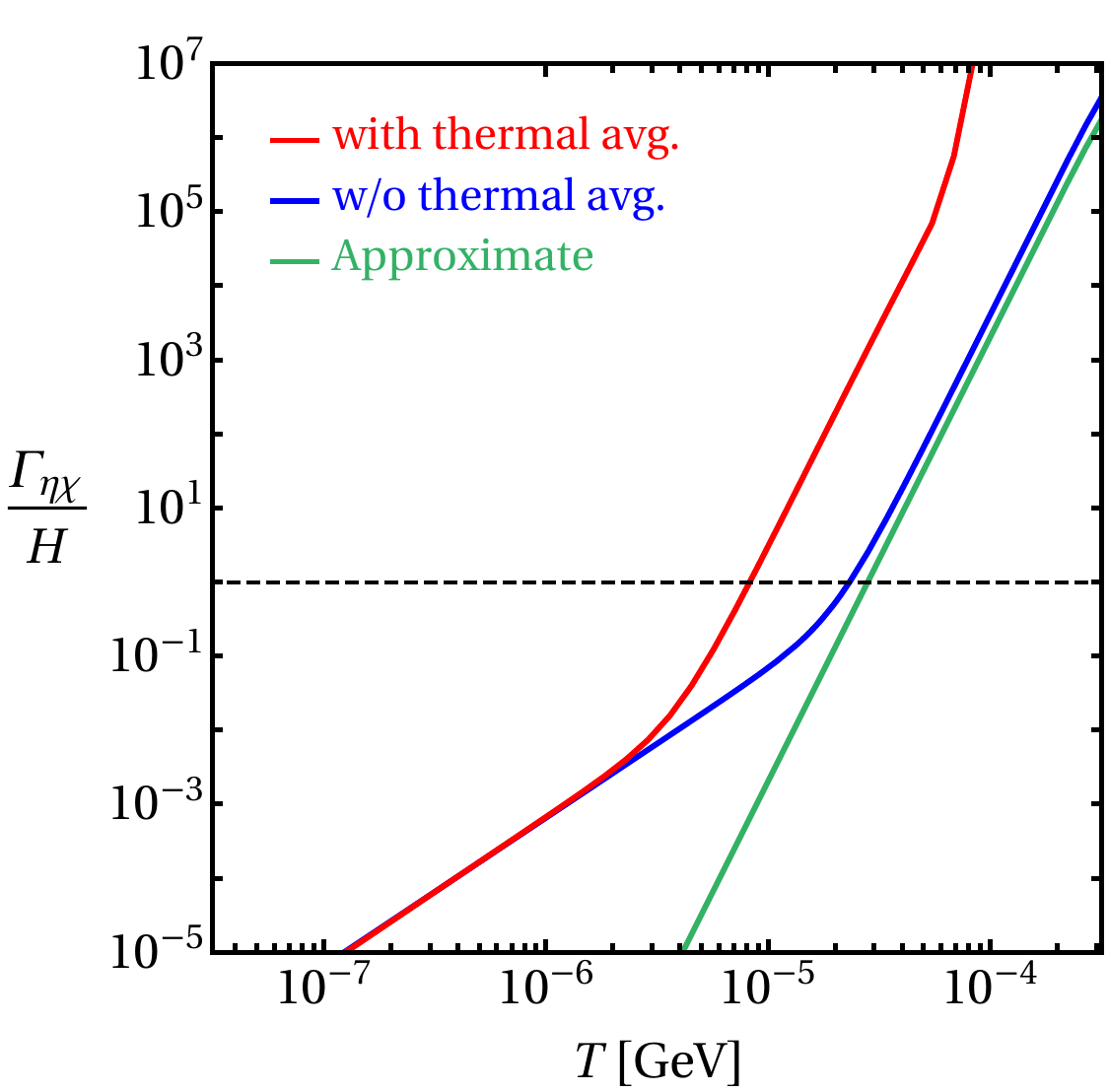}
  \caption{The ratio of the momentum transfer rate $\Gamma_{\eta\chi}$ between $\chi_1$ and $\eta$ to the Hubble parameter $H$. The chosen parameters are $\mchi=1\gev$, $\mrho=1\mev$, $\Delta=1\mev$, $\alpha=10^{-4.5}$.}\label{fig:kinetic_decoupling}
 \end{center}
\end{figure}

In Fig.\,\ref{fig:scattering_v} we show the elastic transfer cross section $\sigma_\text{T}$ as a function of DM velocity $v$ for three different mediator masses. The red curve corresponds to a resonant point in Fig.\,\ref{fig:contour}, and has \mbox{$\sigma_\text{T}\simeq 4\times 10^{-5}\cmg$} at dwarf-scale velocity $v\simeq 10^{-4}$ which is not large enough to produce cores in the DM halos. This is not surprising because indirect detection and CMB experiments put strong constraints on SIDM models with light mediators, and it is difficult to satisfy these constraints along with large self-scattering cross sections to produce core in DM halo\,\cite{Bringmann:2016din}.

However, multilevel SIDM models can generically have interesting scattering phenomenology originating from the inelasticity in the system. DM particles in the ground state can upscatter to the excited state and quickly deexcite to the ground state by emitting a light mediator particle that escapes the halo. As a result, the halo dissipates energy at a certain rate and cools\,\cite{Boddy:2016bbu,Das:2017fyl}. This cooling mechanism can significantly affect the halo dynamics and profile in other parts of the parameter space where self-scattering is large\,\cite{Todoroki:2017kge,Vogelsberger:2018bok}.
%

\subsection{Kinetic decoupling of dark matter}
The scattering between DM $\chi_1$ and $\eta$ keeps the DM in kinetic equilibrium with the relativistic $\eta$ particles longer than the usual. The kinetic decoupling happens roughly when the momentum transfer rate $\Gamma_{\eta\chi}=(T_d/\mchi)n_\chi\sigma_{\eta\chi}$ becomes smaller than the Hubble expansion rate, $\Gamma_{\eta\chi}\lesssim H(T)$. A $t$-channel resonance due to the mass gap $\Delta$ between $\chi_1$ and $\chi_2$ enhances the cross section, and delays the kinetic decoupling of DM further. Such late kinetic decoupling of DM suppresses structure formation at low halo mass scale, and can explain the apparent underabundance of satellite galaxies around the MW\,\cite{Bringmann:2006mu}. It was already pointed out in Ref.\,\cite{Chu:2014lja} that for $\mchi\simeq 1\gev$, $\Delta/\mchi \simeq 10^{-4.5}$ and massless $\eta$ (acting as dark radiation), the kinetic decoupling temperature is $\sim 0.5\kev$ which is essential to solve the missing satellite problem. This temperature decreases further by a factor of a few when a more exact analytic expression for the cross section is used and the thermal distributions of $\chi_1$ and $\eta$ are taken into account, as shown in Fig.\,\ref{fig:kinetic_decoupling}.

However, note that explaining the positron excess requires the DM mass to be \mbox{$\mchi\approx800\gev$} which would increase the decoupling temperature by several orders of magnitude from the $\kev$ scale. Also in the present scenario, the mediator mass $\meta\sim 1\gev$ which ceases to be relativistic much earlier. Thus in the model we present, there is no delayed kinetic decoupling. Of course, one could consider DM with  $M\approx 800\gev$  and a nearly massless $\eta$, but with much smaller $\Delta/\mchi$ to keep the decoupling temperature at keV scale. The Sommerfeld effect phenomenology would change significantly in such a scenario and requires a separate study.

\section{Summary \& Outlook}\label{sec:conclusions}
In this work, we have explored the possibility of explaining the bumps in the galactic positron flux, seen by {AMS-02} and several other experiments, using Sommerfeld-enhanced DM annihilation without overpredicting the DM annihilation signal from dwarf galaxies.

We computed the Sommerfeld-enhanced annihilation rate in a two-level SIDM model, following our previous work in Ref.\,\cite{Das:2016ced}. We show that even though the tree-level annihilation process is $p$-wave suppressed due to the Majorana nature of the DM, the nonperturbative Sommerfeld effect enhances the rate by $\sim\!6$ orders of magnitude. At the same time, the $s$-wave coannihilation is suppressed. This selective Sommerfeld enhancement/suppression was explained to be arising due to a particle exchange symmetry, and occurs only in multilevel DM models.

The dominating $p$-wave annihilation rate has a unique dependence on DM velocity which leads to lower annihilation rate in smaller objects, like dwarf galaxies, but large rate $\sigma v \sim 10^{-24}\cms$ in MW-sized galaxies. This can naturally explain that perhaps the positron flux excess in the cosmic ray observed by AMS-02 and several other experiments is due to the decay of the DM annihilation products into the charged leptons through a higher-dimensional operator. We investigated this possibility and found reasonably good match between the theoretical and observed spectra. 

This is the first work, to the best of our knowledge, interpreting the positron excess as a result of DM annihilation without violating the bounds from dwarf galaxy and CMB observations. One way to test this model is to search for gamma ray lines in, \eg, H.E.S.S. and Fermi-LAT data, originating as final state radiation from the electron and muon. Bremsstrahlung radiation typically shows up as a peak in the gamma ray spectrum around DM mass. However, in the present scenario DM annihilation products $\rho/\eta$ cannot emit photon. Only the $e^\pm/\mu^\pm$, to which the scalars decay into, can emit bremsstrahlung photon. Hence the resulting gamma-ray spectral shape is expected to be different from the usual case. A search for such gamma-ray spectrum can be done from the H.E.S.S. and Fermi-LAT data\,\cite{Abdallah:2018qtu,Li:2019mcx}. Non-observation of such gamma ray emission from dwarf galaxies could serve as an evidence for this model.
We intend to do such analysis in a future work. Moreover, the high energy charged leptons in the decay cascade can also inverse Compton scatter with photons to produce additional signature\,\cite{2010PhRvD..81d3505B}.
The self-scattering cross section in such a DM model turns out to be inadequate to explain the existence of the cored density profiles of the DM halos. 

Even though the positron excess has been seen by both Fermi-LAT and AMS-02, there exists tension between their data in the region $(20-200)\gev$. The reason behind this is not known. Future independent cosmic ray experiments with improved statistics and systematics, and more precise directional flux measurements will pin down the source of the excess.

\section*{Acknowledgments}
We are grateful to Torsten Bringmann, Subhajit Ghosh, and Tuhin Roy for useful discussions. We also thank the anonymous referee for useful suggestions to improve the manuscript. This work was partially funded through a Ramanujan Fellowship of the Dept. of Science and Technology, Government of India, and the Max-Planck-Partnergroup ``Astroparticle Physics'' of the Max-Planck-Gesellschaft awarded to B.D and has received partial support from the European Union's Horizon 2020 research and innovation programme under the Marie-Sklodowska-Curie grant agreement Nos. 674896 and 690575. AD thankfully acknowledges the hospitality provided by Koushik Dutta at SINP during the final stages of this work. We acknowledge use of the {\sc FeynCalc} package~\cite{Shtabovenko:2016sxi}.

\appendix
\section{Nonrelativistic four-Fermi operators}\label{sec:appendix_nr}
In this section, we shall briefly review the two-spinor operators that may arise from the Yukawa interaction between a Dirac fermion $\psi$ of mass $m$ and a scalar $\phi$ or a vector $A_\mu$,
\begin{equation}
	\mathcal{L} \supset y\phi\bar{\psi}\psi\quad \text{or}\quad \mathcal{L} \supset gA^\mu\bar{\psi}\gamma_\mu\psi\,.
\end{equation}
As mentioned before, we shall use the two-component spinor notation for the fermions,
\begin{equation}
\begin{array}{l}
u(\pvec) = \sqrt{\dfrac{E+m}{2E}}\begin{pmatrix}
\xi \\[1ex] \dfrac{\sigmavec\cdot\pvec}{E+m} \xi
\end{pmatrix}\,,\\[6ex]
v(-\pvec) = \sqrt{\dfrac{E+m}{2E}}\begin{pmatrix}
\dfrac{-\sigmavec\cdot\pvec}{E+m} \eta\\[2ex] \eta
\end{pmatrix}\,,\vspace{0.3cm}
\end{array}
\end{equation}
with $E = \sqrt{\pvec^2 + m^2} \simeq m + mv^2/2$ in the nonrelativistic limit. With these definitions, all possible bi-spinor contractions $\bar{v}(-\pvec)\Gamma u(\pvec)$ reduce to--
\begin{equation}
	\begin{array}{lcl}
	\bar{v}(-\pvec)u(\pvec) &=& -\dfrac{1}{m}\eta^\dagger \pvec\cdot\sigmavec\xi\,,\\[2ex]
	\bar{v}(-\pvec)\gamma^0u(\pvec) &=& 0\,,\\[2ex]
	\bar{v}(-\pvec)\gamma^0\gamma^5u(\pvec) &=& \left(1 - \dfrac{\pvec^2}{2m^2}\right)\eta^\dagger\xi \,,\\[2ex]
	\bar{v}(-\pvec)\gamma^iu(\pvec) &=& \eta^\dagger\sigmavec\xi - \dfrac{\pvec^i}{2m^2}\eta^\dagger \pvec\cdot\sigmavec\xi\,,\\[2ex]
	\bar{v}(-\pvec)\gamma^0\gamma^iu(\pvec) &=& -\left(1-\dfrac{\pvec^2}{m^2}\right)\eta^\dagger\sigmavec^i\xi\\[2ex] &&- \dfrac{\pvec^i}{2m^2}\eta^\dagger\pvec\cdot\sigmavec\xi\,,\\[2ex]
	\bar{v}(-\pvec)\gamma^i\gamma^5u(\pvec) &=& \dfrac{i}{m}\eta^\dagger(\pvec\times\sigmavec)^i\xi\,,\\[2ex]
	\bar{v}(-\pvec)\gamma^5u(\pvec) &=& -\eta^\dagger\xi \,.
	\end{array}
\end{equation}
\begin{table}[t]
\begin{center}
\caption{Symmetries of the four-Fermi operators.}\label{tab:nr_operator}
\vspace{0.1cm}
	\begin{tabular}{M{2.6cm}M{3.3cm}}
		\hhline{--}
		Operator & $\spec$\\\hline
		$\mathbf{1}\otimes\mathbf{1}$ & $f(^1S_0)$\\
		$\sigmavec^i\otimes\sigmavec^i$ & $f(^3S_1)$\\
		$v^2\mathbf{1}\otimes\mathbf{1}$ & $h(^1S_0)$\\
		$\vvec'\cdot\vvec \,\mathbf{1}\otimes\mathbf{1}$ & $g(^1P_1)$\\
		$\vvec'\cdot\vvec \,\sigmavec^i\otimes\sigmavec^i$ & $\frac{1}{2}(g(^3P_2) + g(^3P_1))$\\
		$\vvec'\cdot\sigmavec \otimes \vvec\cdot\sigmavec$ & $\frac{1}{3}(g(^3P_0) - g(^3P_2))$\\
		$\vvec\cdot\sigmavec \otimes \vvec'\cdot\sigmavec$ & $\frac{1}{2}(g(^3P_2) - g(^3P_1))$\\
		\hhline{--}
	\end{tabular}
	\end{center}
\end{table} 
The two bi-spinors $f(\pvec)\eta^\dagger\xi$ and $g(\pvec)\eta^\dagger\sigmavec\xi$ are the scalar (spin $s=0$) and vector (spin $s=1$) combinations, respectively. All expressions are approximated to the order $\mathcal{O}((|\pvec|/m)^3)$. In the four-Fermi operators, two bi-spinors are combined together. The orbital angular momentum of such combinations is dictated by the rotational symmetry property of individual operator. A few relevant examples are listed in Table\,\ref{tab:nr_operator}. Here we used the notations: $\vvec = \pvec/m$, $(\xi'^\dagger \eta') (\eta^\dagger \xi) = \mathbf{1} \otimes \mathbf{1}$, $(\xi'^\dagger\sigmavec^i\eta') \cdot (\eta^\dagger\sigmavec^i\xi) = \sigmavec^i \otimes \sigmavec^i$, $(\xi'^\dagger \vvec'\cdot\sigmavec\eta') (\eta^\dagger \vvec \cdot \sigmavec\xi) = \vvec'\cdot\sigmavec \otimes \vvec\cdot\sigmavec$, and $(\xi'^\dagger \vvec \cdot \sigmavec\eta') (\eta^\dagger \vvec'\cdot\sigmavec\xi) = \vvec \cdot \sigmavec \otimes \vvec'\cdot\sigmavec$.

To find the annihilation matrices in Eq.(\ref{eq:ann_matrix}) for a particular process $|a\rangle\to X_AX_B$ where $|a\rangle = \{|\chi_1\chi_1\rangle,\, |\chi_2\chi_2\rangle\}$ for annihilation, and $\{|\chi_1\chi_2\rangle,\, |\chi_2\chi_1\rangle\}$ for coannihilation, we performed the following steps--
\begin{enumerate}
 \item Write down the amplitude for $\Gamma_{ab} = |a\rangle \to X_AX_B \to |b\rangle$ using two-component spinors.
 
 \item Expand it in powers of $|\pvec|/\mchi$ and other small numbers, like $\mrho/\mchi$, $\Delta/\mchi$.
 
 \item Collect the the terms that correspond to an operator $f(\spec)$.

\end{enumerate}

\section{Sommerfeld factor}\label{sec:appendix_sommerfeld}
The effect of the potential is contained in the phase shifts of the wavefunctions of the incoming particles which can be found by solving the \sch{} equation\,\cite{Slatyer:2009vg}. The full wavefunction can be expanded in the partial wave basis as
 \begin{equation}\label{eq:full_wavefunction}
  \Psi(\rvec)_{ij}=\sum_\ell \frac{u_\ell(r)_{ia}}{r}A_{aj}P_\ell(\cos\theta)\,.
 \end{equation}
 Here the expansion coefficients $A_{aj}$ are understood to have an implicit partial wave $\ell$ index. The radial part of the wavefunction obeys the equation
 \begin{equation}
 \label{eq:schrored}
  \Bigg[\bigg(\dfrac{d^2}{dr^2}+k^2-\dfrac{\ell(\ell+1)}{r^2}\bigg)\delta_{ij}
    -2\mu V(r)_{ij}\Bigg]u_\ell(r)_{jk} = 0\,.
 \end{equation}
 The asymptotic form of the regular solutions of the above equation is
 \begin{equation}
  u_\ell(r)_{ij}\overset{r\to\infty}{=}N_{ij}\sin\left(k_ir-\frac{\ell\pi}{2}+s_{ij}\right)\,,
 \end{equation}
 which we compare with the large radius asymptotic form of the total wavefunction $\Psi({\bf r})$ in Eq.(\ref{eq:full_wavefunction}) to yield
 \begin{equation}
  A_{ij}=i^\ell(2\ell+1)\frac{\left[M^{-1}\right]_{ij}}{k_i}
 \end{equation}
 where we used $M_{ij}=N_{ij}e^{-is_{ij}}$. Here $s_{ij}$ denote the phase shifts due to the scattering in respective channel. In the last two equations, no summation is implied over any repeated index. To compute the Sommerfeld factor, we need the value and the radial derivative of the wavefunction at the origin. For this purpose, we series expand the reduced wavefunction  $u_\ell(r)$ as
 \begin{equation}\label{eq:smallr}
  u_\ell(r)=\frac{1}{(\ell+1)!}\frac{d^{\ell+1}u_\ell(r)}{dr^{\ell+1}}\bigg|_{r=0}r^{\ell+1}+\ldots
 \end{equation}
 for all four elements. To use this form of the wavefunction near the origin, one needs to know the phase shift matrix $M$. Instead we use the $r$-independence of the Wronskian of the linearly independent solutions of the \sch{} equation. We shall compute the Wronskian both at $r\to 0$ and $r\to\infty$ limits and compare them to obtain the $M$ matrix in terms of the amplitude of the wavefunctions at infinity. The irregular set of solutions to Eq.(\ref{eq:schrored}), defined as $v(r)_{ij}$, have the asymptotic behaviour
 \begin{equation}\label{eq:irr_sol}
  v(r)_{ij}\overset{r\to 0}{=}\delta_{ij}r^{-\ell},\quad v(r)_{ij}\overset{r\to\infty}{=}T^\dagger_{ij}e^{-ik_ir}.
 \end{equation}
 The Wronskian is defined as
 \begin{equation}\label{eq:wron}
  W_\ell(r)_{ij}\equiv v^\dagger_\ell(r)_{ik}u'_\ell(r)_{kj} - v^{\dagger}_{\ell}{'}(r)_{ik}u_\ell(r)_{kj}\,.
 \end{equation}
 Using the asymptotic forms of the solutions, we find
  \vbox{
  \begin{align}
   W_\ell(r)_{ij}\overset{r\to 0}{=}\dfrac{2\ell+1}{(\ell+1)!}\dfrac{d^{\ell+1}u_\ell(r)}{dr^{\ell+1}}\bigg|_{r=0}\,,\\[3.5ex] 
   W_\ell(r)_{ij}\overset{r\to\infty}{=}i^\ell\sum_a k_a T_{ia} M_{aj}\,.
  \end{align}}
 Now equating these two quantities yields the phase shift matrix $M$ in terms of the large-$r$ amplitude $T$, which used in the full solution of Eq.(\ref{eq:sch}) gives the value of the wavefunction at origin in terms of $T$,
 \begin{equation}
  \Psil(0)_{ij}=(2\ell-1)!!\,i^{-\ell}\frac{T^\dagger_{ij}}{k^\ell_i}\,.
 \end{equation}
 Using this expression for the wavefunction at the origin in the definition of Sommerfeld factor yields Eq.(\ref{eq:sommerfeld}). The $T$ matrix is computed by solving the \sch{} equation numerically. A brief algorithm is provided below, following Ref.\,\cite{Slatyer:2009vg}.
 \begin{enumerate}
  \item Eq.(\ref{eq:schrored}) is solved in between $r=r_0$ and $r_f$. The point $r_0$ is chosen such that the centrifugal term dominates over the potential and Eq.(\ref{eq:smallr}) is valid. The initial conditions are as follows
  \begin{equation}
  \begin{array}{l}
   u_\ell(r_0)_{ij}=\dfrac{r_0^{\ell+1}}{2\ell+1}\,\delta_{ij},\\[2.3ex] 
   u'_\ell(r_0)_{ij}=\dfrac{(\ell+1)r_0^\ell}{2\ell+1}\,\delta_{ij}\,.
  \end{array}
  \end{equation}
  With this choice of the boundary conditions, the Wronskian turns out to be exactly identity from Eq.(\ref{eq:wron}).

  \item At $r=r_f$, the Wronskian is written as below, using Eq.(\ref{eq:irr_sol}) and (\ref{eq:wron}).
  \begin{equation}
   \begin{array}{rcl}
    (W_\ell)_{ij} &=& T_{ia}\left[u'_\ell(r_f)_{aj}-ik_iu_\ell(r_f)_{aj}\right]e^{ik_ir_f}\\
    &=& \delta_{ij}\,.
   \end{array}
  \end{equation}
  
  \item The $T$ matrix obtained by inverting the $B$ matrix which is defined below
  \begin{equation}\label{eq:B_matrix}
  \begin{array}{l}
   T=B^{-1},\\[2.3ex] 
   B_{ij}\equiv \left[u'_\ell(r_f)_{ij}-ik_iu_\ell(r_f)_{ij}\right]e^{ik_ir_f}.
  \end{array}
  \end{equation}

 \end{enumerate}
 
 This procedure works well when the heavier annihilation channel $|\chi_2\chi_2\rangle$ is kinematically open. When that is not the case, the wavefunctions $\langle\chi_2\chi_2|\chi_1\chi_1\rangle$ or $\langle\chi_2\chi_2|\chi_2\chi_2\rangle$ are exponentially growing/decaying. The matrix inversion in Eq.(\ref{eq:B_matrix}) with those solutions becomes difficult and gives rise to numerical instabilities. To mitigate this problem, we have followed the modified variable phase method as prescribed in Ref.\cite{Beneke:2014gja}.
 
 \ad{
 \section{Equivalent one-level system}\label{sec:appendix_onelevel}
 The two-level \sch{} equations in Eq.(\ref{eq:sch}) are restated below for readers' convenience.
 \begin{equation}
 \label{eq:sch_app}
 \left[\mathcal{D}_{ab} + V_{ab}(r)\right]\left(\ul(r)\right)_{bc}=0\,.
\end{equation}
Note that the indices $a,b,c$ run over the two-body states $|\chi_1\chi_1\rangle, |\chi_2\chi_2\rangle$ for annihilation, and $|\chi_1\chi_2\rangle, |\chi_2\chi_1\rangle$ for coannihilation. Let us explicitly write down the first equation (11-component) from Eq.(\ref{eq:sch_app}),
 \begin{equation}\label{eq:sch_11}
\left(\mathcal{D}_{11} + V_{11}\right)u_{11} + V_{12} u_{21}=0\,,
\end{equation}
where we have suppressed all other indices except that of two-body states to avoid clutter. Now using the particle exchange symmetry in the annihilation subspace,
\begin{equation}
	u_{21} = \langle \chi_2\chi_2|\chi_1\chi_1\rangle \approx (-1)^{\ell+s} \langle \chi_1\chi_1|\chi_1\chi_1\rangle = (-1)^{\ell+s} u_{11}\,.
\end{equation}
This equation relates the transition amplitude $u_{21}$ to $u_{11}$ through the factor $(-1)^{\ell+s}$. Using it to substitute $u_{21}$ in Eq.(\ref{eq:sch_11}) gives
\begin{equation}
\left[\mathcal{D}_{11} + V_\mathrm{eff}\right] u_{11}=0\,,
\end{equation}
 which is the equivalent one-level \sch{} equation with the effective potential given by
 \begin{equation}
 V_\mathrm{eff} = V_{11} + (-1)^{\ell+s} V_{12}\,.
 \end{equation}
 A similar derivation is applicable for coannihilation as well.
}

\bibliographystyle{apsrev4-1}
\bibliography{dmdr}

\end{document}